\documentclass[twocolumn,trackchanges]{aastex701}

\usepackage{amsmath,amssymb}
\usepackage{subcaption}
\usepackage{comment}

\usepackage[normalem]{ulem}
\usepackage{hyperref}

\begin{document}

\newcommand{\note}[1]{\textcolor{magenta}{#1}}
\newcommand{\D}{\text{d}}
\newcommand{\modot}{\text{M}_\odot}
\newcommand{\rodot}{\text{R}_\odot}
\newcommand{\Hz}{\text{Hz}}
\newcommand{\pbh}{\text{PBH}}
\newcommand{\WD}{\text{WD}}
\newcommand{\pc}{\text{pc}}
\newcommand{\Mpc}{\text{Mpc}}
\newcommand{\Kpc}{\text{Kpc}}
\newcommand{\km}{\text{km}}
\newcommand{\seconds}{\text{s}}
\newcommand{\hr}{\text{hr}}
\newcommand{\yr}{\text{ yr}}
\newcommand{\secinv}{\text{s}^{-1}}
\newcommand{\p}{\texttt{p}}
\newcommand{\bhpt}{\textsc{BHPTNRSur1dq1e4}}
\newcommand{\few}{\textsc{FEW}}

\newif\ifredmarkup
\redmarkupfalse

\title{
Detectability of Eccentric Intermediate Mass-Ratio Inspiral in the Decihertz Band
}

\author[0009-0006-3310-481X]{Devesh Giri}
\affiliation{Indian Institute of Science Education and Research, Pashan, Pune 411008, India}
\email{deveshgiri.edu@gmail.com}

\author[0000-0001-7035-1373]{Sajal Mukherjee}
\affiliation{Department of Physics, Birla Institute of Technology and Science -- Pilani, Rajasthan 333031, India}
\email{sajal.mukherjee@pilani.bits-pilani.ac.in}

\author[0000-0001-7394-0755]{Apratim Ganguly}
\affiliation{Inter-University Centre for Astronomy and Astrophysics (IUCAA), Post Bag 4, Ganeshkhind, Pune 411007, India}
\email{apratim@iucaa.in}

\author[0000-0002-0800-4626]{Sanjit Mitra}
\affiliation{Inter-University Centre for Astronomy and Astrophysics (IUCAA), Post Bag 4, Ganeshkhind, Pune 411007, India}
\email{sanjit@iucaa.in}

\begin{abstract}
The proposed decihertz gravitational-wave (GW) band $(\sim 10^{-2}-1\,\Hz)$ will observe the late inspiral of intermediate mass-ratio inspirals (IMRIs), binaries with mass ratios $10^2 \lesssim q \lesssim 10^4$ that assemble dynamically in dense stellar environments and are expected to retain measurable eccentricity. Search and parameter-estimation pipelines are predominantly built around circular templates, and we test this assumption. Using black-hole perturbation theory waveforms from \textsc{FastEMRIWaveforms} for a representative IMRI ($M=10^4\,\modot$, $q=100$, observed for its final $10\,\hr$), we compute fitting factors between eccentric signals and circular templates, maximizing over the template's total mass and mass ratio with a differential-evolution search, for the B-DECIGO, LGWA-Si, and IndIGO-D noise curves and a flat PSD. The mismatch grows with initial eccentricity $e_0$ and crosses the standard $3\%$ threshold at $e_0\simeq3\times10^{-2}$, well within the range allowed by the waveform model and expected from dynamical formation. The best-fitting circular templates also bias the total mass and mass ratio by several percent at $e_0=0.1$. Combining the mismatch with the signal-to-noise-dependent distinguishability criterion, we find that for a source at $1\,\mathrm{Gpc}$, eccentricities as small as $e_0\simeq5\times10^{-4}$ (B-DECIGO and IndIGO-D) and $5\times10^{-3}$ (LGWA-Si) are resolvable, one to two orders of magnitude below where detection efficiency degrades. The loss of match arises from the excitation of radial Teukolsky harmonics, which spread power into sidebands that circular templates cannot represent, and from the slower orbital decay of eccentric binaries at fixed initial semi-latus rectum. Robust detection and parameter estimation of IMRIs with DECIGO, LGWA, and IndIGO-D will require eccentric waveform models and template banks.
\end{abstract}
\section{Introduction}\label{sec:intro}

The proposed decihertz gravitational-wave (GW) band ($\sim10^{-2}$--$1\,\mathrm{Hz}$), bridging the millihertz band targeted by the Laser Interferometer Space Antenna (LISA) and the $\gtrsim10\,\mathrm{Hz}$ band probed by ground-based detectors, is a central science driver for proposed missions such as the DECi-hertz Interferometer Gravitational wave Observatory (DECIGO), Indian Initiative in Gravitational-wave Observations-D (IndIGO-D)~\citep{IndIGOD:2024repo,Sharma:2026tff}, the Lunar Gravitational-Wave Antenna (LGWA), and the Big Bang Observer (BBO)~\citep{Seto:2001qf,Amaro-Seoane:2018gbb,Harry:2006fi,Yagi:2011wg,Ajith:2024mie}. An important class of target sources is intermediate mass-ratio inspirals (IMRIs) -- binaries with mass ratios $10^{2}\lesssim q\lesssim10^{4}$~\citep{Nguyen:2026jft, Haster:2016ewz, Amaro-Seoane:2018gbb} that occupy the regime between the comparable-mass binaries routinely observed by the LIGO-Virgo-KAGRA~\citep{LIGOScientific:2016aoc,LIGOScientific:2018mvr,LIGOScientific:2020ibl,KAGRA:2021vkt} and the extreme mass-ratio inspirals (EMRIs) targeted by LISA~\citep{LISA:2017pwj}. IMRIs are expected to assemble dynamically, in dense stellar clusters, hierarchical triples, and AGN-disks~\citep{Miller:2001ez,Antonini:2012ad,Fragione:2018vty,Leigh:2017wff,Secunda:2020mhd}, formation channels that leave the resulting binary with non-negligible orbital eccentricity~\citep{Peters:1964zz,Wen:2002km,OLeary:2008myb,Samsing:2017xmd}. 

Intermediate-mass black holes are the missing link between stellar-mass and supermassive black holes, and how they form and grow remains poorly understood. The orbital eccentricity of an IMRI is one of the few observables that carries the answer. A binary that evolves in isolation circularizes through gravitational-wave emission long before it reaches the decihertz band \citep{Peters:1964zz}. However, a binary assembled dynamically through the channels above can enter the band with eccentricity that has not yet been radiated away. Measuring the eccentricity of an IMRI can therefore reveal which of these channels produced it. This depends on how small an eccentricity a detector can distinguish from a circular orbit, and on whether the searches used to find these sources preserve that signature. We address both questions here for a representative decihertz IMRI.

Eccentricity imprints distinctive signatures on GW signals. In contrast to circular binaries, whose radiation is dominated by a single harmonic of the orbital frequency, eccentric systems emit across a spectrum of harmonics, producing frequency sidebands separated by the radial orbital frequency~\citep{Peters:1963ux,1977ApJ...216..610T}. Even for circular binaries, the radiation may have significantly higher-harmonic content if the mass asymmetry is large~\citep{Giri:2025pkx, Islam:2021zee}. Circular binaries, however, have no radial-mode content, whereas the rich radial-mode content of eccentric binaries results in characteristic amplitude and phase modulations in the time domain, as well as a multi-track structure in time-frequency representations. For EMRIs, these effects have been extensively studied using black hole (BH) perturbation theory and Teukolsky-based waveform models~\citep{Drasco:2003ky,Barack:2003fp,Drasco:2005is}, where they are known to play a central role in parameter estimation.

IMRIs can retain measurable eccentricity through the entire observable inspiral~\citep{Mancieri:2025cmx, Strokov:2023kmo, Amaro-Seoane:2018gbb}, which for a decihertz IMRI spans hours to days and $10^{3}$--$10^{4}$ radians of accumulated orbital phase~\citep{Laxman:2026xts}. This matters because the traditional search and parameter-estimation pipelines for compact binary-like sources are predominantly built, implicitly or explicitly, around circular or quasi-circular templates. For long-lived, high-phase-accumulation decihertz IMRIs, even a small departure from circularity can accumulate into a search-relevant dephasing~\citep{Amaro-Seoane:2018gbb, Martel:1999tm, Wang:2023tle}. Whether circular templates remain adequate for these sources is therefore the central question. The answer determines what a decihertz IMRI search can detect and how well it can characterize the source.

To generate the gravitational waveform, we implement the \textsc{FastEMRIWaveforms} (\textsc{FEW}) framework~\citep{Chua:2017ujo,Katz:2021yft,Chapman-Bird:2025xtd}, which is based on BH perturbation theory. We further use this waveform to compute the fitting factor between eccentric IMRI signals and the best available circular template, as a function of the initial eccentricity $e_0$, for a representative decihertz IMRI observed for its final $10\,\mathrm{hr}$ of evolution before merger. Our results show that circular templates recover a rapidly shrinking fraction of the optimal signal-to-noise ratio (SNR) as $e_0$ grows, crossing the standard $3\%$ mismatch threshold at $e_0\simeq3\times10^{-2}$. This eccentricity is small compared to both the range permitted by the waveform model across the decihertz IMRI parameter space and also the eccentricities plausible for dynamically formed binaries. Two effects drive the loss of match: first, eccentricity moves power into radial sidebands, and a circular template has no radial harmonics to hold it. Second, the orbital phase of an eccentric binary evolves differently from that of a circular one, starting at the same separation. Refitting the template’s mass and mass ratio may adjust the difference, but at the price of biasing both significantly. Through this work, we also discuss the minimum eccentricity required to have any statistically resolvable imprint on the recovered source parameters as a function of luminosity distance.

The rest of the paper is organized as follows. Section~\ref{sec:waveforms} describes the waveform model, and the representative system, and Section~\ref{sec:ff} defines the fitting-factor formalism. We present the central results in Section~\ref{sec:results} and conclude the paper with a brief discussion in Section~\ref{sec:discussion}. Appendix~\ref{app:morphology} presents the waveform morphology and phase evolution, and Appendix~\ref{app:match} gives the explicit form of the match for multipolar waveforms.
\section{Waveform Modeling and Flux-Based Eccentricity Limits}\label{sec:waveforms} 
\begin{figure*}[t]
    \centering
    \begin{subfigure}[b]{0.48\textwidth}
    \centering
    \includegraphics[width=\textwidth]{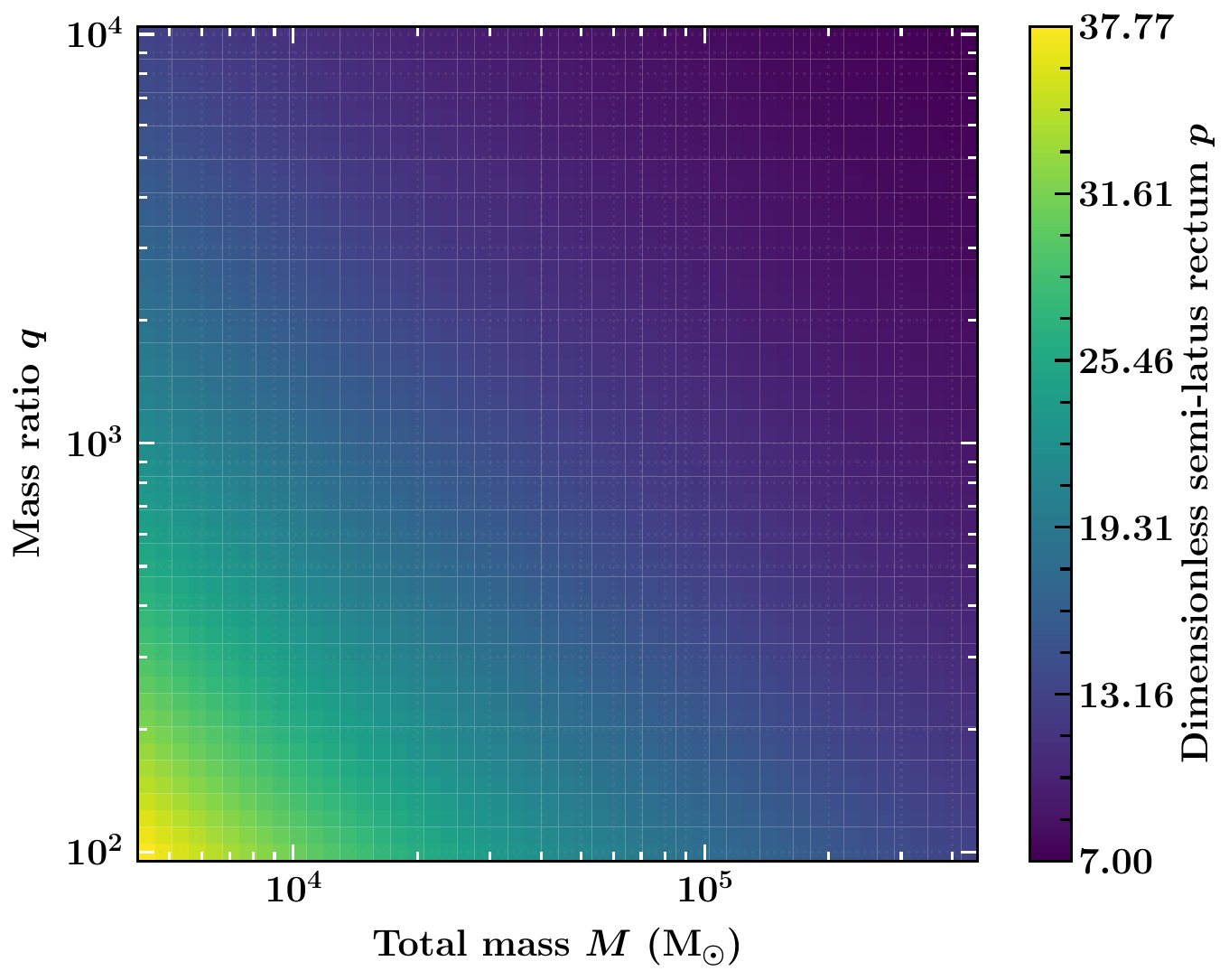}
    \caption{}
    \label{fig:p0grid_a}
    \end{subfigure}
    \hfill
    \begin{subfigure}[b]{0.48\textwidth}
    \centering
    \includegraphics[width=\textwidth]{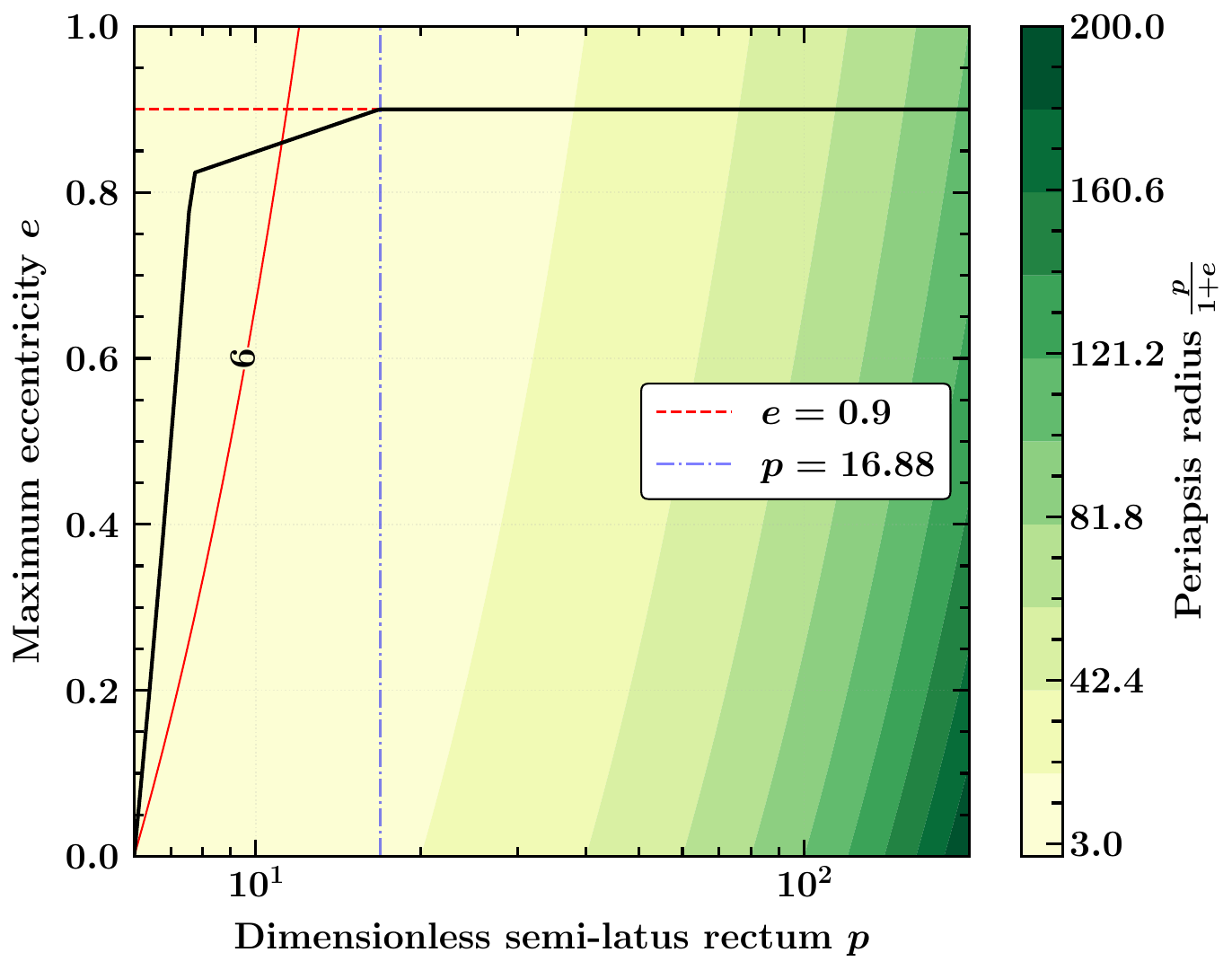}
    \caption{}
    \label{fig:p0grid_b}
    \end{subfigure}
    \caption{(a) Initial semi-latus rectum $p_0$ required for $T_{\rm obs}=24\,\mathrm{hr}$ across the decihertz $(M,q)$ parameter space. (b) Maximum allowed eccentricity $e_{\rm max}$ as a function of $p$ for the non-spinning \texttt{KerrEccEqFlux} model, saturating at $e=0.9$ for $p\gtrsim16.88$. The black curve follows the Schwarzschild separatrix, $p=6+2e$, at low $p$ (up to $e\approx0.8$). At larger $p$ the model’s eccentricity taper sets the limit. The thin red contour marks where the periapsis radius reaches the circular-orbit ISCO radius, $p/(1+e)=6$.}
    \label{fig:p0grid}
\end{figure*}

We generate all inspiral trajectories and waveforms with \few, which implements adiabatic BH perturbation theory for eccentric, equatorial Kerr inspirals by solving the Teukolsky equation~\citep{Drasco:2003ky,Barack:2003fp,Drasco:2005is,Burke:2023lno}. We restrict to non-spinning ($a=0$), equatorial, prograde orbits, described at each instant by the dimensionless semi-latus rectum (in units of $M$) and eccentricity, $\{p(t),e(t)\}$. In addition, we adopt the \texttt{FastKerrEccentricEquatorialFlux} waveform model with the \texttt{KerrEccEqFlux} trajectory ~\citep{Chapman-Bird:2025xtd}. This model is valid only up to a maximum eccentricity $e_{\rm max}(p)$ set by the extent of its pre-computed flux grids, as shown in Figure \ref{fig:p0grid}. These flux grids are computed \textit{offline} in a discrete grid, and then interpolated \textit{on the fly} during integrating the trajectory. This is because a direct numerical solution of the Teukolsky equations at each point along the trajectory would be computationally too expensive~\citep{Katz:2021yft}. For high eccentricity and small $p$ values, the numerical solution becomes unreliable, and it is this truncation that defines $e_{\rm max}(p)$.

The azimuthal, polar, and radial motion of the secondary is tracked by three fundamental orbital phases $\{\Phi_\varphi(t),\Phi_\theta(t),\Phi_r(t)\}$. For the Schwarzschild equatorial case, the polar and azimuthal frequencies are degenerate, $\hat\Omega_\theta=\hat\Omega_\varphi$, so $\Phi_\theta(t)=\Phi_\varphi(t)$. The radial frequency $\Omega_r$ remains well defined even for circular orbits, and $\Phi_r(t)$ is integrated for all $0\le e<1$. The waveform is a sum over modes labeled by the spherical-harmonic index $\ell$, azimuthal harmonic $m$, and radial harmonic $n$ ($\ell\ge2$, $|m|\le\ell$, $n\in\mathbb{Z}$; the ranges supported by FEW are given below). We define the dimensionless mode frequencies as $\hat\omega_{mn}=m\hat\Omega_\varphi+n\hat\Omega_r$, and the cumulative phase of each mode as $\Phi_{mn}(t)=m\Phi_\varphi(t)+n\Phi_r(t)$. Each mode's amplitude, $H_{\ell mn}$, is determined by the homogeneous Teukolsky amplitude $ \hat{Z}^{\infty}_{\ell mn}$. The strain at null infinity reads as
\begin{align}
    h(t) &\equiv h_+(t)-ih_\times(t) \notag \\
    &= \frac{\mu}{D_L}\sum_{\ell mn}H_{\ell mn}(t,\theta,\varphi)\,e^{-i\Phi_{mn}(t)},
    \label{eq:strain}
\end{align}
where $q\equiv m_1/m_2$, $M=m_1+m_2$, $\mu=m_1m_2/M$, $D_L$ is the luminosity distance, and $(\theta,\varphi)$ are the observer's viewing angles. The inspiral terminates at the separatrix, which for Schwarzschild is $p_{\rm sep}=6+2e$ and reduces to the familiar ISCO, $p_{\rm sep}=6$, for circular orbits. Here we set the initial conditions with {\texttt{get\_p\_at\_t}}, which uses Brent's root-finding {~\citep{Brent1971AlgorithmsFF, Brent1971AnAW}} to determine the initial semi-latus rectum {$p_0$} corresponding to a specified observation time {$T_{\rm obs}$}. 

For circular orbits ($e=0$) only $n=0$ modes survive, with the dominant contribution from the quadrupole $(\ell,m,n)=(2,2,0)$ mode. The sub-dominant $(\ell,m,0)$ modes can also be excited by mass asymmetry alone~\citep{Giri:2025pkx}. For eccentric orbits, $n\neq 0$ modes acquire amplitudes that scale as powers of $e$, generating sidebands spaced by $\Omega_r$ around each azimuthal harmonic, as discussed later in Section~\ref{sec:results_mechanism}. Within \few, the supported mode indices are $\ell\in[2,10]$, $|m|\le\ell$, $k=0$ (the polar harmonic index, which vanishes for equatorial orbits), and $|n|\le55$. The computational cost is controlled by a mode-selection threshold $\varepsilon$, which sorts modes by power and keeps them until the retained power reaches a fraction $1-\varepsilon$ of the total. Here we adopt $\varepsilon=10^{-3}$ throughout the paper.

The trajectory ODE is defined over $p\in(6.002,200)$, $e\le e_{\rm max}(p)$. To find the largest initial eccentricity the model allows across the decihertz IMRI parameter space, we focus on IMRIs with mass ratios $10^{2}\le q\le10^{4}$ whose inspiral reaches the ISCO within the decihertz band, $0.01\,\mathrm{Hz}\le f_{\rm ISCO}\le1\,\mathrm{Hz}$. Using GW frequency at ISCO in Schwarzschild, $f_{\rm ISCO}=c^{3}/(6^{3/2}\pi GM)\approx4.40\times10^{3}\,\mathrm{Hz}\,(M_\odot/M)$ \citep{Bardeen:1972fi}, this range maps to the total-mass range $4.4\times10^{3}\,M_\odot\le M\le4.4\times10^{5}\,M_\odot$ used throughout this paper. Because the inspiral ends at the separatrix, fixing the observation duration $T_{\rm obs}$ fixes the initial semi-latus rectum $p_0(M,q)$ of the circular inspiral. We use this value as the common initial separation for all eccentricities, so that circular and eccentric signals enter the band at the same orbital separation\footnote{We fix $p_0$ rather than the semi-major axis $a=p/(1-e^2)$, since $p$ is the natural variable of the flux grids and of the separatrix.}. At fixed $p$, the orbital decay of an eccentric binary is slower than that of a circular one, so the eccentric orbit remains at larger $p$ throughout, and only the circular signal reaches the separatrix at $t=T_{\rm obs}$. The eccentric signals end before plunge; for $e_0=0.1$ the window closes at $p\simeq8.08$, compared with $p_{\rm sep}\simeq6.03$. The signal starts at $(p_0,e_0)$, where $e_0$ is the eccentricity at this initial point. For each $(M,q)$ in this range, we ask how large an $e_0$ the model allows at the corresponding $p_0$.

The model's $e_{\rm max}(p)$ saturates at $0.9$ for $p\gtrsim 16.88$ and falls steeply at smaller $p$ (Figure \ref{fig:p0grid_b}). We combine it with $p_0(M,q)$ for $T_{\rm obs}=24\,\mathrm{hr}$ over a $20\times20$ logarithmic grid in $(M,q)$, as shown in Figure \ref{fig:p0grid_a}. This gives maximum initial eccentricities of $e_{\rm max}\simeq 0.9$ at the lowest masses and mass ratios, falling to $\simeq0.5$ at the highest, so the eccentricities studied here lie well inside the model’s validity domain. The value of $p_0$ decreases with both $M$ and $q$, from $p_0\simeq 37.8 $ at the low-$(M,q)$ corner to $p_0\simeq 7.0$ at the high-$(M,q)$ corner. Far from the ISCO ($p_0\gg6$) this follows the Newtonian quadrupole merger-time scaling \citep{Peters:1964zz}
\begin{align}
    p_0 \simeq \left(\frac{256\,c^{3}T_{\rm obs}}{5\,GMq}\right)^{1/4} \propto (Mq)^{-1/4}. \label{eq:p0_scaling}
\end{align}
The scaling is only approximate here. Note that Equation~\ref{eq:p0_scaling} gives $p_0\simeq24.7$ instead of $25.04$ for the representative system below, and fails near the ISCO, where it gives $p_0\simeq3.8$ (inside the ISCO) at the high-$(M,q)$ corner. The values of $p_0$ used in this work are therefore obtained with the full relativistic adiabatic flux.

For the detailed waveform analysis, we adopt a representative binary near the low-mass, low-$q$ corner of the parameter space, observed for a shorter duration:
\begin{align}
    M=10^{4}\,M_\odot,\quad q=100,\quad T_{\rm obs}=10\,\mathrm{hr},
\end{align}
for which $p_0\approx25.04$ and the model supports eccentricities up to $e_{\rm max}\simeq0.9$. Despite spanning only the final $10\,\mathrm{hr}$ before merger, this window accumulates enough phase for eccentricity to produce considerable observable effects (Appendix~\ref{app:morphology}). We consider four representative initial eccentricities, $e_0\in\{0,10^{-4},10^{-2},0.1\}$ ({circular, small, moderate, and large, respectively}), spanning the transition from near-circular to appreciably eccentric inspirals while remaining well within the model's validity range.
\section{Fitting-Factor Calculations}\label{sec:ff}

For the fitting-factor calculations we fix a representative system, $M=10^{4}\,\modot$, $q=100$, observed for its final $T_{\rm obs}=10\,\mathrm{hr}$ before merger, where the initial semi-latus rectum is $p_0\approx25.04$ (Section \ref{sec:waveforms}). We evaluate the fitting factor on a dense grid of 50 logarithmically spaced initial eccentricities spanning $e_0\in[10^{-3},10^{-1}]$, and additionally generate full waveforms at four representative values of Section~\ref{sec:waveforms}, $e_0\in\{0,10^{-4},10^{-2},0.1\}$, for the morphological comparisons in Appendix~\ref{app:morphology}. All other binary parameters are held fixed, so that any difference between signals is attributable to $e_0$ alone. Every recovery template is the corresponding circular ($e=0$) waveform from the same model, with its total mass and mass ratio left free. Throughout this paper, $M$ denotes the detector-frame (redshifted) total mass, so that the waveforms computed here are the observed ones at the quoted luminosity distance; the mass ratio $q$ is frame-invariant. The corresponding source-frame total mass is $M/(1+z)$, a distinction that becomes relevant in Section~\ref{sec:results_emin}.

\begin{figure}
    \centering
    \includegraphics[width=\columnwidth]{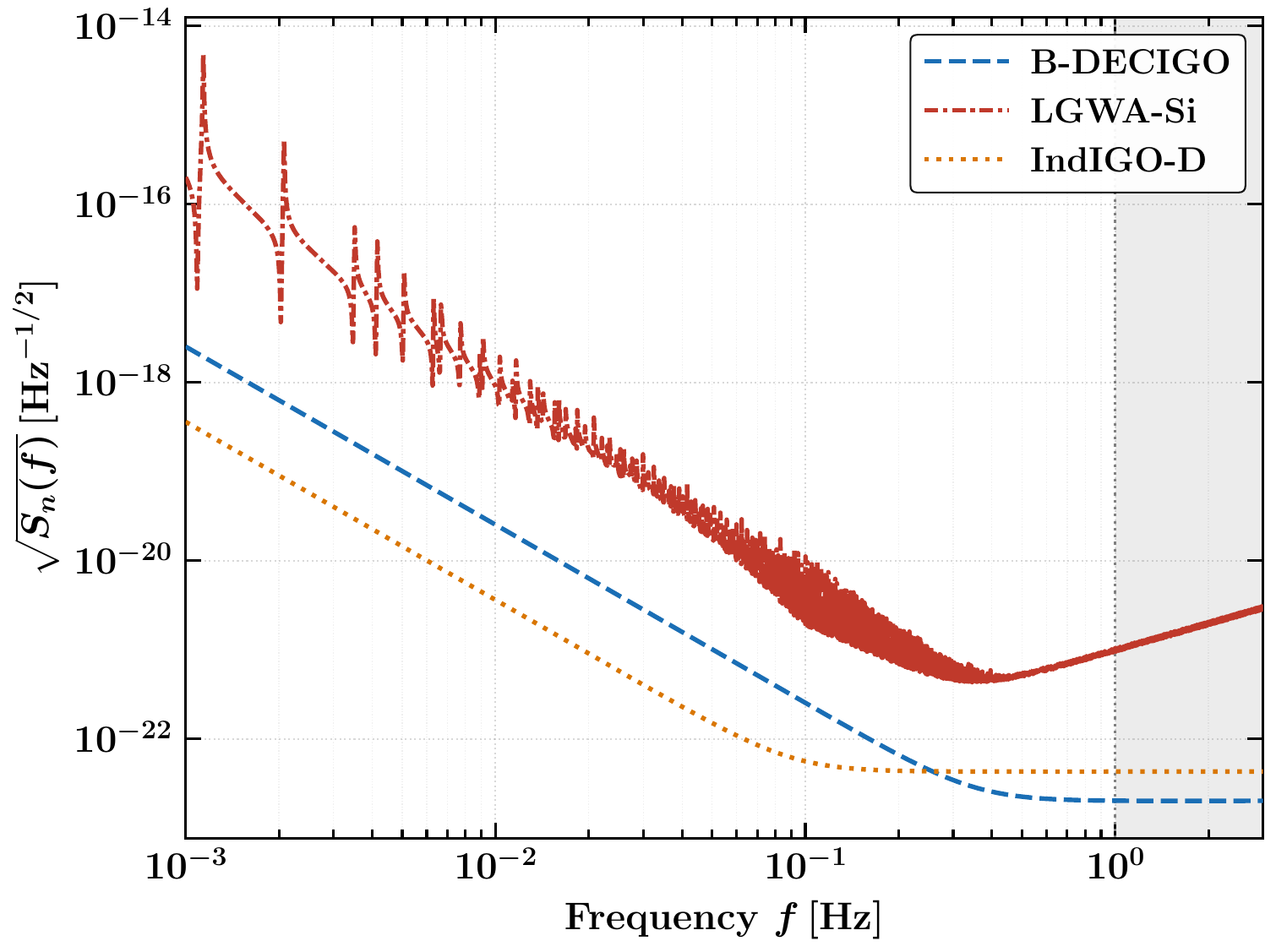}
    \caption{Amplitude spectral densities $\sqrt{S_n(f)}$ of the three detector noise curves used in this work: B-DECIGO \citep{Isoyama:2018rjb}, the LGWA-Si baseline \citep{Ajith:2024mie}, and IndIGO-D in its S2 configuration \citep{IndIGOD:2024repo, Sharma:2026tff}. All overlaps are evaluated over $[10^{-3},1]\,\mathrm{Hz}$; the shaded region above $1\,\mathrm{Hz}$ is excluded.}
    \label{fig:psd}
\end{figure}

\begin{figure}
    \centering
    \includegraphics[width=\columnwidth]{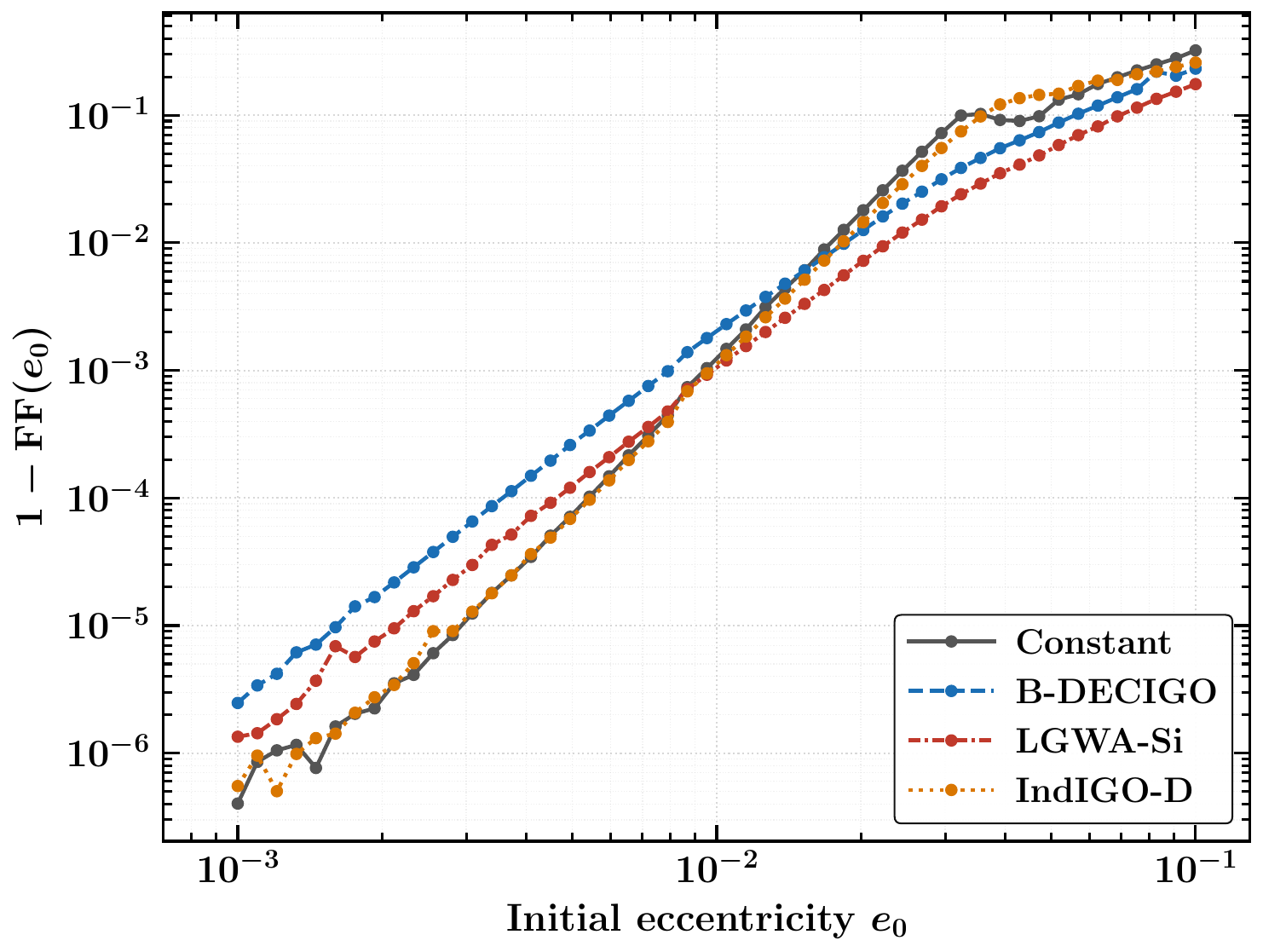}
    \caption{Mismatch, $1-\mathrm{FF}$, between eccentric IMRI signals and the best-fitting circular template as a function of the initial eccentricity $e_0$, for the representative system ($M=10^{4}\,M_\odot$, $q=100$, $T_{\rm obs}=10\,\mathrm{hr}$) and four noise curves. The standard $3\%$-mismatch search threshold is crossed at $e_0\simeq3\times10^{-2}$ for the three realistic detector configurations. The curve labelled ``Constant'' is the flat-PSD case.}
    \label{fig:mismatch}
\end{figure}

For two frequency-domain waveforms $\tilde h_1(f)$ and $\tilde h_2(f)$ we adopt the noise-weighted inner product $(a|b)=4\,\mathrm{Re}\int_{f_{\rm low}}^{f_{\rm high}} \tilde a(f)\tilde b^{*}(f)/S_n(f)\,\mathrm{d}f$ \citep{Cutler:1994ys}, the overlap $\mathcal{O}(h_1,h_2)=(h_1|h_2)/\sqrt{(h_1|h_1)(h_2|h_2)}$, and the match $\mathcal{M}=\max_{t,\Phi}\mathcal{O}$, maximized over the relative time and a single overall phase (see Appendix~\ref{app:match}). The fitting factor,
\begin{align}
    \mathrm{FF} = \max_{m_1,m_2}\mathcal{M}, \qquad \mathcal{MM}=1-\mathrm{FF},\label{eq:ff}
\end{align}
additionally maximizes over the template's component masses, with the signal drawn from the eccentric family and the template restricted to circular inspirals, following standard fitting-factor methodology \citep{Apostolatos:1995pj,Owen:1998dk,damouriyersathya,Varma:2014jxa}. $\mathcal{MM}$ is the fractional loss of matched-filter SNR incurred by searching for an eccentric IMRI with a circular template. We maximize Equation~\ref{eq:ff} over the template's $(\log M,\log q)$ with the differential-evolution algorithm \citep{Storn:1997de} as implemented in \textsc{SciPy} \citep{scipy}. For each trial template, the maximization over relative time and phase is carried out by the \texttt{optimized\_match} function of \textsc{PyCBC}~\citep{Usman:2015kfa}. We repeat the full calculation for four noise curves -- B-DECIGO, the LGWA-Si baseline configuration, the IndIGO-D S2 configuration, and a flat PSD included for reference -- so that our conclusions do not depend on a single assumed sensitivity curve (Figure~\ref{fig:mismatch}). The B-DECIGO noise curve is taken from \cite{Isoyama:2018rjb}, the LGWA-Si baseline curve from \cite{Ajith:2024mie}, the IndIGO-D curve is the S2 (moderate-sensitivity) model of \cite{IndIGOD:2024repo, Sharma:2026tff}, and the flat PSD is a constant $S_n(f)$, whose normalization drops out of the match. The three detector noise curves are compared in Figure~\ref{fig:psd}. All overlaps are computed on the complex strain $h=h_+-ih_\times$ over $[f_{\rm low},f_{\rm high}]=[10^{-3},1]\,\mathrm{Hz}$ on a frequency grid of spacing $\Delta f=10^{-5}\,\mathrm{Hz}$, for the fixed viewing angles $(\theta,\varphi)=(\pi/3,0)$.
\section{Results}\label{sec:results}

\begin{figure*}[ht!]
    \centering
    \begin{subfigure}[b]{0.48\textwidth}
    \centering
    \includegraphics[width=\textwidth]{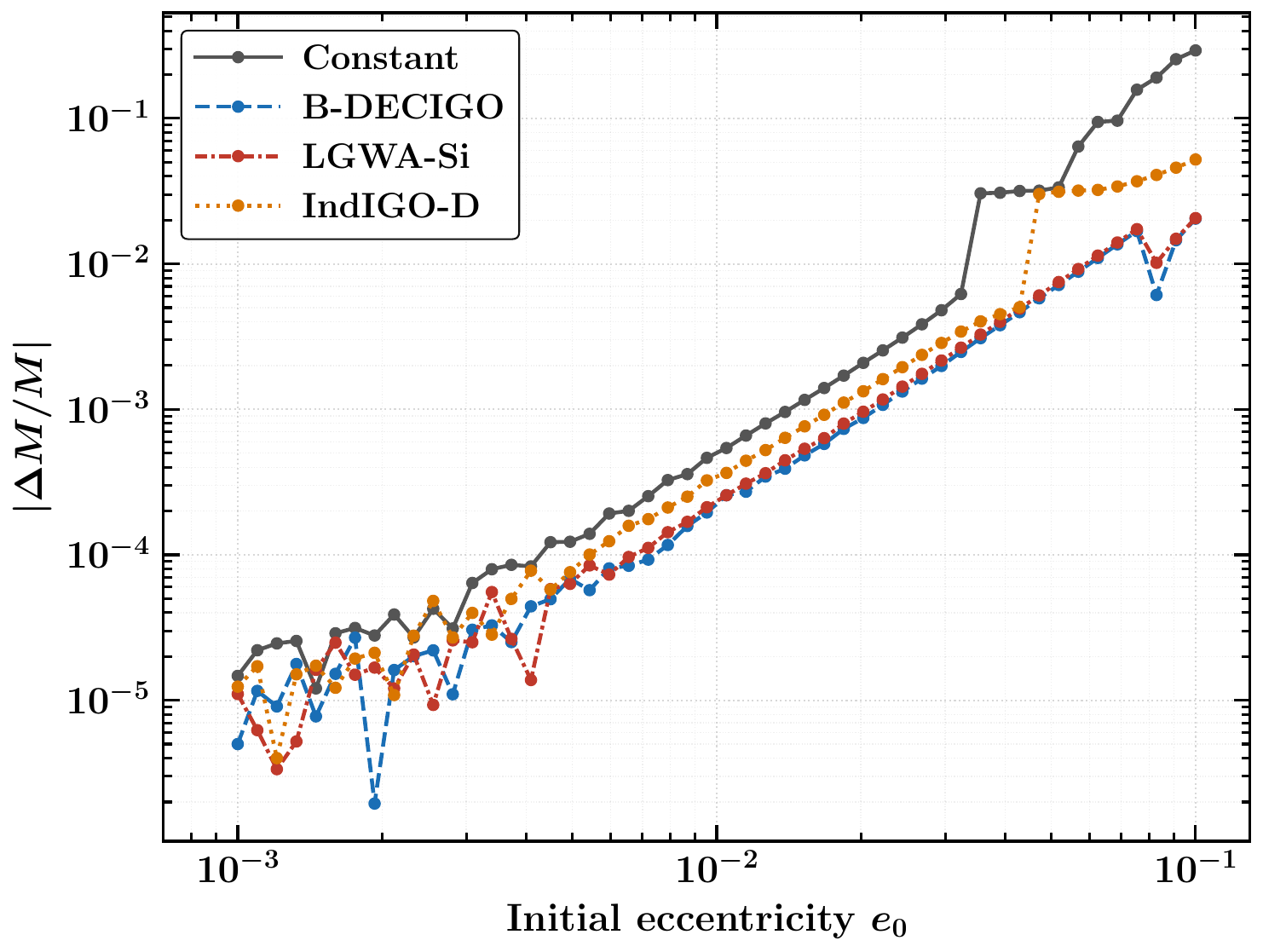}
    \caption{}
    \label{fig:bias_M}
    \end{subfigure}
    \hfill
    \begin{subfigure}[b]{0.48\textwidth}
    \centering
    \includegraphics[width=\textwidth]{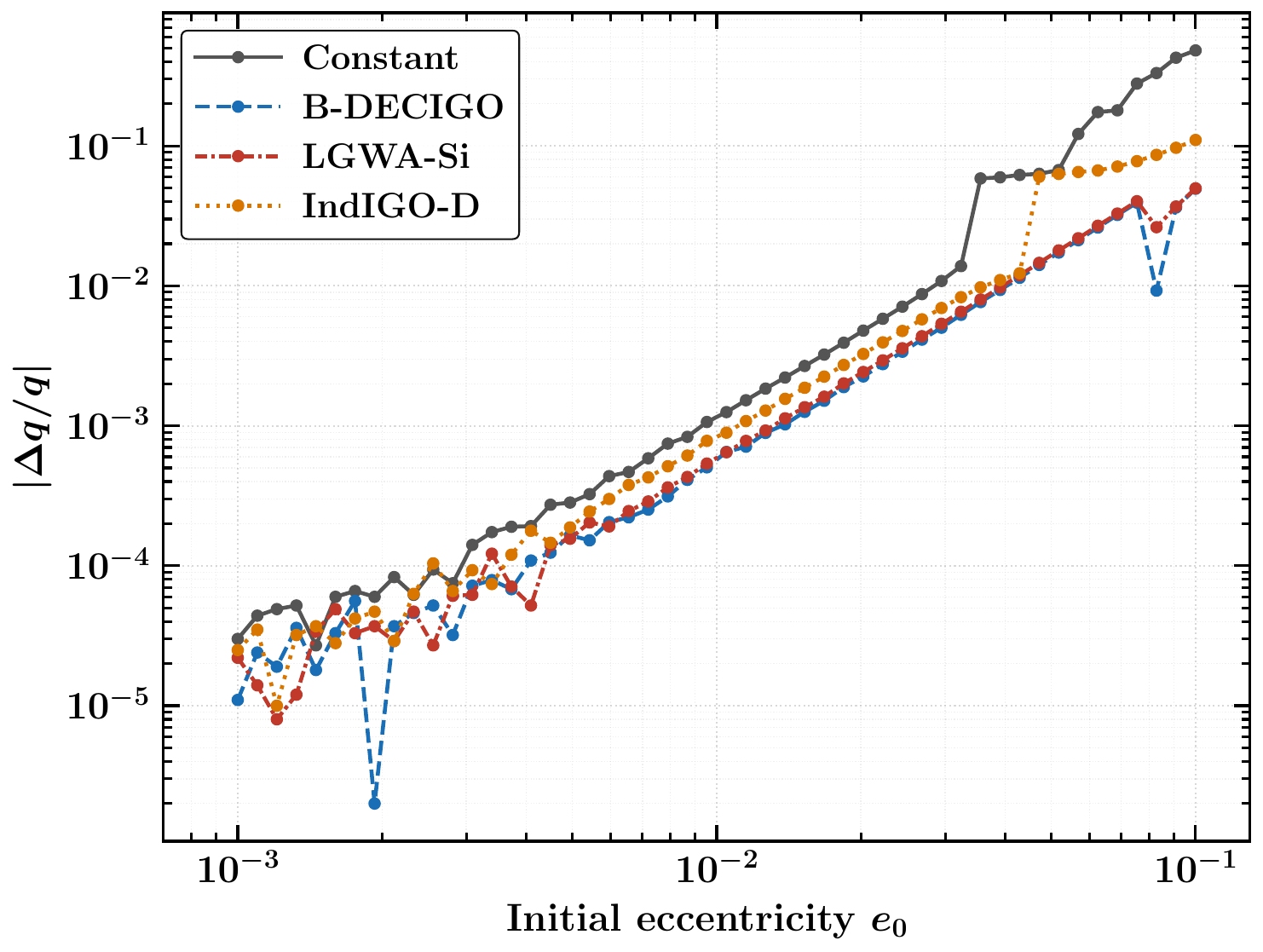}
    \caption{}
    \label{fig:bias_q}
    \end{subfigure}
    \caption{Fractional bias in the recovered (a) total mass, $|\Delta M|/M$, and (b) mass ratio, $|\Delta q|/q$, for the best-fitting circular template, as a function of the initial eccentricity $e_0$, for the same four noise curves as in Figure~\ref{fig:mismatch}. Biases grow systematically with $e_0$ for every detector PSD: a circular template compensates for the slower inspiral of an eccentric signal by drifting to different masses, not by reproducing the missing radial harmonics.}
    \label{fig:bias}
\end{figure*}

\subsection{Mismatch and Parameter Bias of Circular Templates}\label{sec:results_mismatch}

Figure~\ref{fig:mismatch} shows the mismatch $\mathcal{MM}=1-\mathrm{FF}$ between eccentric IMRI signals and the best-fitting circular template, as a function of $e_0$, for all four noise curves. The mismatch rises essentially monotonically with $e_0$, by nearly five orders of magnitude across the range studied: from $\mathcal{MM}\sim10^{-6}$ at $e_0=10^{-3}$, where the fractional SNR loss is negligible, to $\mathcal{MM}\gtrsim10^{-1}$ at $e_0=0.1$. As shown in Section~\ref{sec:results_emin}, however, a negligible SNR loss does not imply that the eccentricity is negligible for inference. The trend is detector independent to a good approximation, since the noise PSD reweights the frequency integral in the inner product rather than altering the underlying waveform disagreement. Across the whole range, the four curves agree to within a factor of a few, compared with the five orders of magnitude over which each of them varies.

Adopting the standard search criterion $\mathrm{FF}\gtrsim0.97$ ($\mathcal{MM}\lesssim3\%$), the threshold is crossed at $e_0\simeq3\times10^{-2}$ for the three realistic noise curves considered here, from $e_0\simeq2.4\times10^{-2}$ (IndIGO-D) to $\simeq3.7\times10^{-2}$ (LGWA-Si). This result is worth highlighting on its own merit -- an initial eccentricity of a few percent, deep within the bound-orbit domain, well within the range expected from any dynamical IMRI formation channels and well within the range the waveform model permits throughout the decihertz parameter space ({Section~\ref{sec:waveforms}}), is already enough for a circular-template search to lose an appreciable fraction of the recoverable SNR. Beyond this eccentricity, circular-template searches suffer a potentially significant loss in detection efficiency for eccentric IMRIs.

We shall clarify what the $3\%$ criterion does and does not imply, since it is a detection-oriented convention rather than a fundamental limit. It originates in template-bank construction, where $\mathrm{FF}\gtrsim0.97$ corresponds to a fractional SNR loss of $\lesssim3\%$ and hence, for sources distributed uniformly in volume, to a loss of $\lesssim10\%$ in detection rate~\citep{Owen:1995tm,Owen:1998dk}. By construction, it is independent of the SNR of any individual source. The criterion that is relevant at the level of parameter estimation is generally stricter: two waveforms yield statistically distinguishable posteriors once $(\delta h|\delta h)\gtrsim D$, that is, once $\mathcal{MM}\gtrsim D/(2\rho^{2})$, with $\rho$ the optimal SNR and $D$ the number of parameters varied~\citep{Flanagan:1997kp,Lindblom:2008cm,Baird:2012cu}. We take $D=2$, matching the $(M,q)$ maximization of Equation~\ref{eq:ff}. A full Bayesian analysis sampling more parameters would relax the threshold by an $\mathcal{O}(1)$ factor. Since the threshold is dominated by the $\rho^{-2}$ scaling, this does not affect the conclusions drawn below. The injected SNRs for our representative system are listed in Table~\ref{tab:snr}. For B-DECIGO ($\rho\simeq1.6\times10^{3}$), LGWA-Si ($\rho\simeq1.2\times10^{2}$), and IndIGO-D ($\rho\simeq3.2\times10^3$) the distinguishability threshold lies at $\mathcal{MM}\sim4\times10^{-7}$, $\sim7\times10^{-5}$, and $\sim1\times10^{-7}$ respectively, roughly five (B-DECIGO), between two and three (LGWA-Si), and between five and six (IndIGO-D) orders of magnitude below the detection criterion. So eccentricity begins to bias astrophysical inference at eccentricities far smaller than $e_0\simeq3\times10^{-2}$ (as can be seen in Figure~\ref{fig:bias}, the mass and mass-ratio biases are already growing steadily at $e_0\sim10^{-3}$).

Maximizing the match $\mathcal{M}$ over the template's total mass and mass ratio partially compensates for the missing eccentric structure. But it does so by shifting the recovered parameters away from their true values rather than by genuinely matching the signal. Figure~\ref{fig:bias} shows the resulting fractional biases in the recovered total mass, $|\Delta M|/M$, and mass ratio, $|\Delta q|/q$, for the best-fitting circular template. Both are negligible for near-circular binaries but grow systematically with $e_0$, mirroring the mismatch trend of Figure~\ref{fig:mismatch}. For B-DECIGO and LGWA-Si, the fractional biases reach $\mathcal{O}(1\%)$ by $e_0\sim0.05$. The flat-PSD case, included for reference, overestimates the bias by up to an order of magnitude for $e_0\gtrsim0.039$, since it weights frequency bands that a realistic detector PSD suppresses. For IndIGO-D, up to $e_0\simeq0.043$ its biases exceed those of B-DECIGO and LGWA-Si by at most a factor of $\simeq2$; at $e_0\simeq0.047$ the best fit moves to a different circular template, with $|\Delta M|/M\simeq3\%$ and $|\Delta q|/q\simeq6\%$, much like the jump of the flat-PSD case at $e_0\simeq0.039$. At $e_0=0.1$ its biases ($\simeq5\%$ in $M$ and $\simeq11\%$ in $q$) remain well below the flat-PSD values ($\simeq29\%$ and $\simeq48\%$), consistent with a noise floor that is nearly white above $\simeq0.1\,\mathrm{Hz}$ but rises steeply below it (Figure~\ref{fig:psd}). In every case, the recovered circular template compensates for the slower inspiral of the eccentric signal by shifting to different masses, but this adjustment cannot reproduce the additional radial harmonics or the associated amplitude modulation. Therefore, eccentricity biases the astrophysical inference even when the circular-template search itself still marginally succeeds ($e_0\lesssim3\times10^{-2}$).

Table~\ref{tab:snr} lists the optimal SNR, $\rho=\sqrt{(h|h)}$, of the injected eccentric signals for the representative system of Section \ref{sec:ff} at $D_L=1\,\mathrm{Gpc}$. It is computed on the same frequency grid and over the same band $[10^{-3},1]\,\mathrm{Hz}$ used for the fitting-factor calculation. We have computed the optimal SNR for three realistic detector noise curves. Values are shown for a representative subset of the eccentricity grid, whereas the full scan comprises 50 logarithmically spaced points in $e_0\in[10^{-3},10^{-1}]$.
\begin{table}[h]
    \centering
    \caption{Optimal SNR of the injected eccentric IMRI signals ($M=10^{4}\,M_\odot$, $q=100$, $p_0=25.04$, $T_{\rm obs}=10\,\mathrm{hr}$, $D_L=1\,\mathrm{Gpc}$) for three decihertz detector configurations.}
    \begin{tabular}{lccc}
         \hline\hline
         $e_0$ & B-DECIGO & LGWA-Si & IndIGO-D \\
         \hline
         $0.0010$ & $1645.3$ & $117.00$ & $3153.5$ \\
         $0.0049$ & $1644.9$ & $117.01$ & $3153.3$ \\
         $0.0105$ & $1643.2$ & $117.04$ & $3153.0$ \\
         $0.0202$ & $1630.1$ & $116.70$ & $3150.8$ \\
         $0.0518$ & $1546.4$ & $112.69$ & $3135.7$ \\
         $0.1000$ & $1378.0$ & $103.46$ & $3096.8$ \\
         \hline
    \end{tabular}
    \label{tab:snr}
\end{table}
Two features are worth noting. First, the three configurations differ by an order of magnitude and a few factors in SNR, which is the origin of the differing distinguishability thresholds quoted in Section~\ref{sec:results_mismatch} and of the vertical offset between the three curves in Figure~\ref{fig:emin_dL}. Adopting a nominal single-detector detection threshold of $\rho=8$, the near circular SNRs correspond to horizon distances of $\simeq206\,\mathrm{Gpc}$ for B-DECIGO, $\simeq15\,\mathrm{Gpc}$ for LGWA-Si, and $\simeq394\,\mathrm{Gpc}$ for IndIGO-D for this system. Second, the SNR itself degrades with increasing eccentricity, by $\simeq16\%$ (B-DECIGO), $\simeq12\%$ (LGWA-Si), and only $\simeq2\%$ (IndIGO-D) across the scan, as power moves from the dominant $(\ell,m,0)$ tracks into radial sidebands, some of which fall outside the most sensitive part of the band. The much smaller loss for IndIGO-D is consistent with its noise floor being nearly white above $\simeq0.1\,\mathrm{Hz}$ (Figure~\ref{fig:psd}), so that redistributing power in frequency within the band changes its SNR little. This loss is intrinsic to the signal and is distinct from, and additional to, the template-mismatch loss quantified in Section~\ref{sec:results_mismatch}.

\subsection{Minimum Distinguishable Eccentricity}\label{sec:results_emin}

\begin{figure}
    \centering
    \includegraphics[width=\columnwidth]{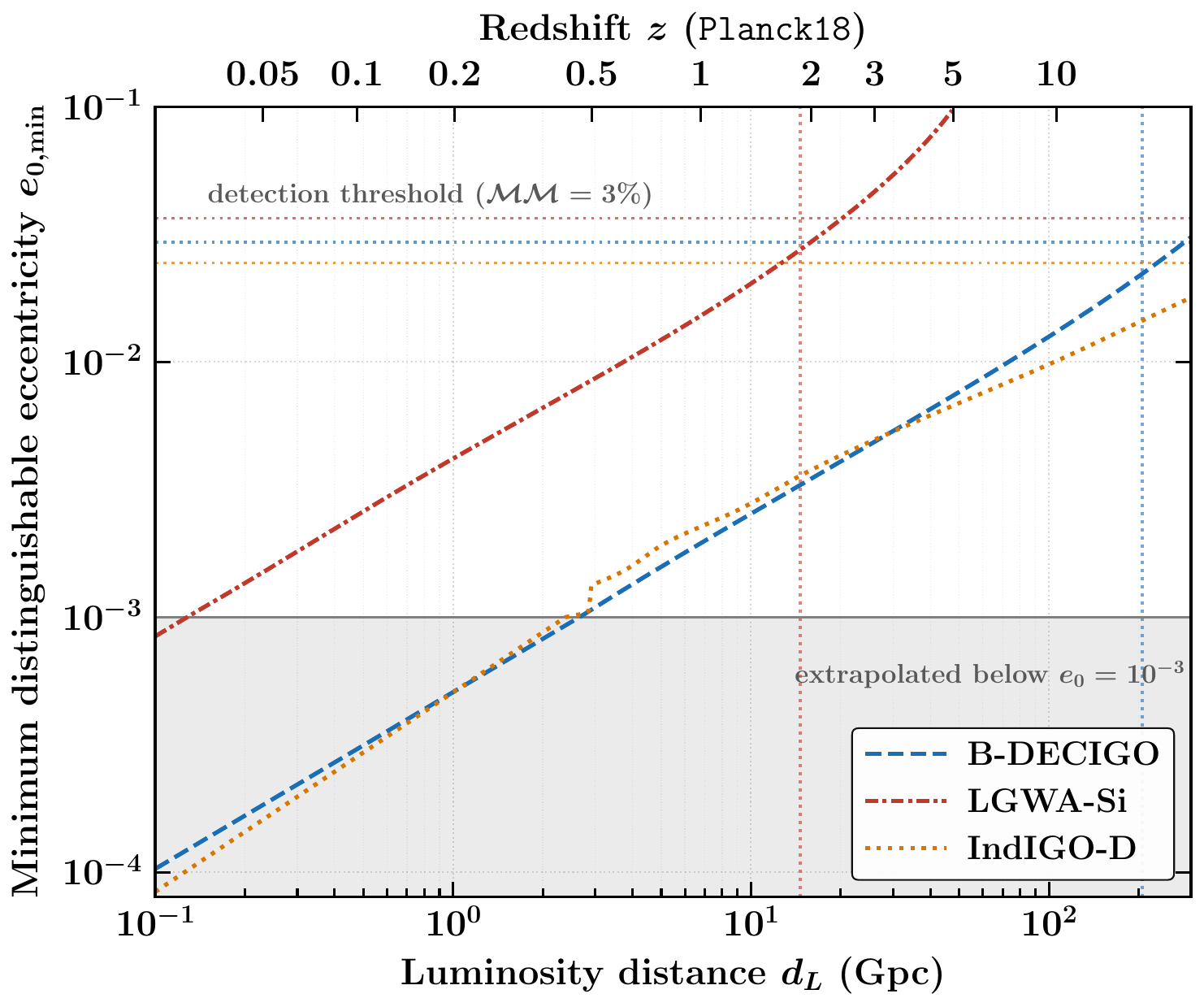}
    \caption{Minimum eccentricity $e_{0,\mathrm{min}}$ at which an eccentric decihertz IMRI $(M\sim10^4\,M_\odot)$ becomes statistically distinguishable from the best-fitting circular template, as a function of luminosity distance, obtained by inverting the measured $\mathcal{MM}(e_0)$ curve of Figure~\ref{fig:mismatch} at the threshold $\mathcal{MM}=D/(2\rho^{2})$ with $\rho\propto d_L^{-1}$. Dotted horizontal lines mark the detection-oriented $3\%$ mismatch criterion for each noise curve; vertical dotted lines mark the corresponding horizon distances for this system; the IndIGO-D horizon ($\simeq394\,\mathrm{Gpc}$) lies beyond the plotted range. The shaded band below $e_0=10^{-3}$ indicates where $e_{0,\mathrm{min}}$ falls outside the simulated eccentricity range and the fitted power law has been extrapolated. The upper axis converts $d_L$ to redshift using \texttt{Planck18}; masses are detector-frame throughout, so the corresponding source-frame total mass is $M_{\rm src}=M/(1+z)$.}
    \label{fig:emin_dL}
\end{figure}

The two criteria of Section~\ref{sec:results_mismatch} can be combined into a single statement with direct observational content. The mismatch curve $\mathcal{MM}(e_0)$ of Figure~\ref{fig:mismatch} is a property of the waveforms and the noise curve alone and is independent of distance. The distinguishability threshold $\mathcal{MM}\gtrsim D/(2\rho^{2})$ depends on distance only through the SNR. Because the masses used throughout are detector-frame (redshifted) masses, the observed waveform at luminosity distance $d_L$ is obtained from the one computed here purely by the amplitude rescaling $\rho \propto d_L^{-1}$\footnote{Under a redshift the binary rescales as $M\to(1+z)M$, $t\to(1+z)t$, $f\to f/(1+z)$, while the dimensionless orbital elements $p$ and $e$ are invariant, so $e_0$ denotes the same physical eccentricity at every distance.}. Inverting the measured $\mathcal{MM}(e_0)$ at the distance-scaled threshold therefore yields the smallest eccentricity that leaves a statistically distinguishable imprint on the recovered parameters,
\begin{align}
    e_{0,\mathrm{min}}(d_L) \;=\; \mathcal{MM}^{-1}
    \!\left[\frac{D\,d_L^{2}}{2\,\rho_{\rm ref}^{2}\,d_{\rm ref}^{2}}\right],
    \label{eq:emin}
\end{align}
with $\rho_{\rm ref}$ the SNR at the reference distance $d_{\rm ref}=1\,\mathrm{Gpc}$ (Table~\ref{tab:snr}). We evaluate Equation~\ref{eq:emin} by fitting a cubic spline to $\log\mathcal{MM}$ versus $\log e_0$, continued below the simulated range with the power law fitted to its low-eccentricity end. We use $\rho_{\rm ref}$ values of $1645.3$ (B-DECIGO), $117$ (LGWA-Si), and $3153.5$ (IndIGO-D).

Figure~\ref{fig:emin_dL} shows that at $1\,\mathrm{Gpc}$ a B-DECIGO-like detector distinguishes eccentricities as small as $e_{0,\mathrm{min}}\sim$ $\mathcal{O}(10^{-4})$, and LGWA-Si $e_{0,\mathrm{min}}\sim$ $\mathcal{O}(10^{-3})$. IndIGO-D reaches essentially the same value as B-DECIGO, $e_{0,\mathrm{min}}\simeq5\times10^{-4}$, despite its twice larger SNR: its mismatch is a factor of $\sim4$ lower at $e_0=10^{-3}$ (Figure~\ref{fig:mismatch}) and falls less steeply towards smaller $e_0$ (power-law index $\simeq2.6$ versus $\simeq2.9$), which offsets its $\simeq3.7$ times stricter distinguishability threshold. Two limits should be kept in mind when reading Figure~\ref{fig:emin_dL}. First, $e_{0,\mathrm{min}}$ is meaningful only where the source is detectable at all: for the system considered here the horizon distance is $\simeq15\,\mathrm{Gpc}$ for LGWA-Si, $\simeq206\,\mathrm{Gpc}$ for B-DECIGO (vertical dotted lines), and $\simeq394\,\mathrm{Gpc}$ for IndIGO-D (beyond the plotted range), and beyond these the curves are indicative rather than observationally realizable because of the size of the observational universe. Second, where $e_{0,\mathrm{min}}$ falls below $10^{-3}$ (shaded band), the value follows from extrapolating the fitted power law beyond the simulated eccentricity grid. These values should be read as extrapolated estimates of the eccentricity that can safely be neglected, rather than as measured thresholds. Both the measured and extrapolated values are conservative, since the mode selection omits part of the sideband power at small $e_0$ (Appendix~\ref{app:match_def}). Within these limits, Figure~\ref{fig:emin_dL} sets the eccentricity resolution floor for decihertz IMRI parameter estimation of this representative system across the astrophysically accessible volume. We recall that the source-frame total mass corresponding to our detector-frame $M=10^{4}\,\modot$ is $M/(1+z)$, ranging from $\simeq10^{4}\,\modot$ locally to $\mathcal{O}(10^{3})\,\modot$ at the highest redshifts shown.

\subsection{Physical Origin: Radial Harmonic Excitation}\label{sec:results_mechanism}

\begin{figure*}
    \centering
    \includegraphics[width=\textwidth]{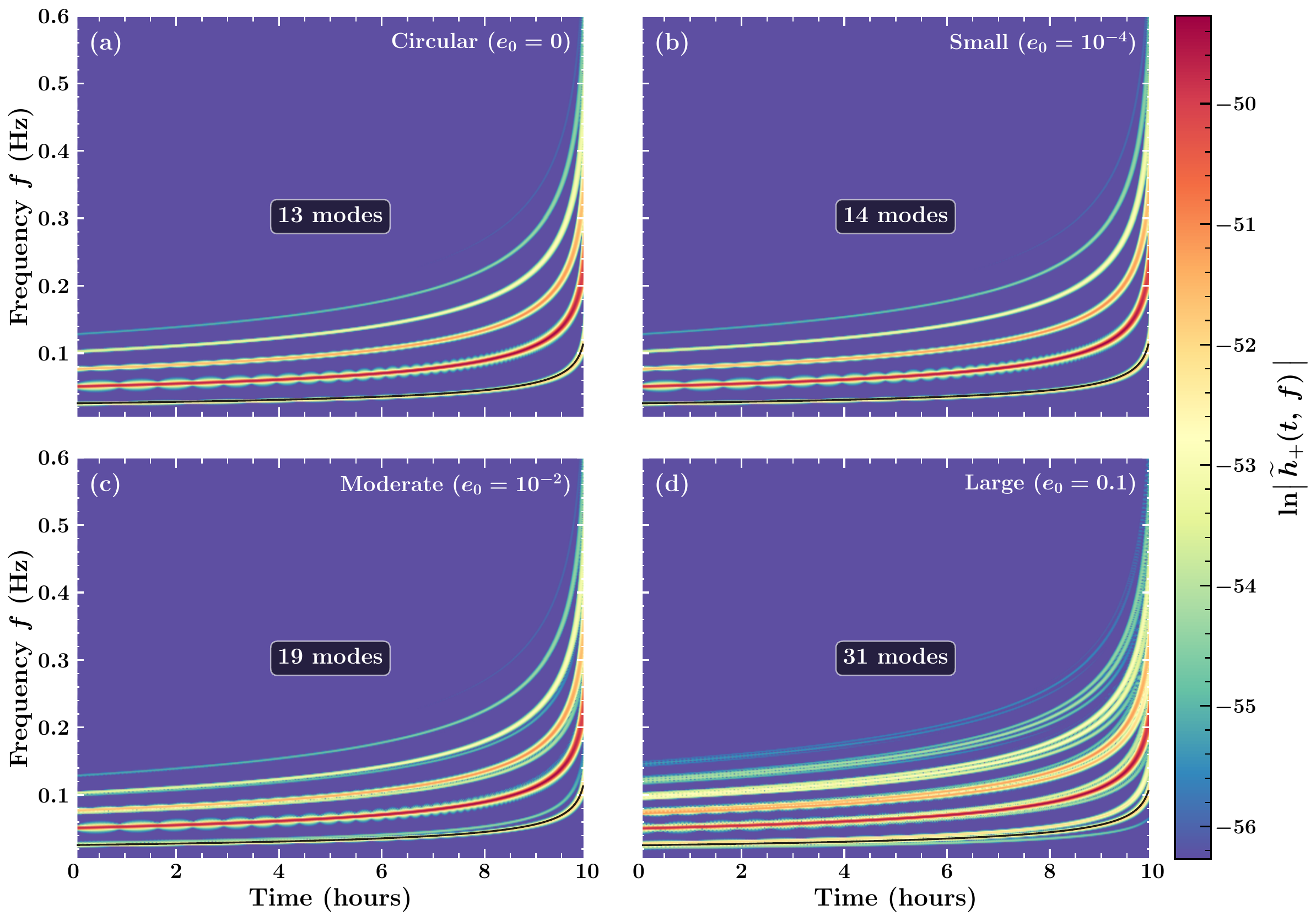}
    \caption{Time-frequency spectrograms of $h_+$ for the representative system at four initial eccentricities, colored by $\ln|\tilde h_+(t,f)|$ on a common scale. The excitation of radial sidebands around each pre-existing $(\ell,m,0)$ track, growing in prominence from panel (a) to (d), is one of the two contributions to the mismatch and parameter bias of Figure~\ref{fig:mismatch} and Figure~\ref{fig:bias}: a circular template contains only $n=0$ modes and cannot represent this sideband power at any choice of mass or mass ratio. Mode amplitudes are shown in Figure~\ref{fig:modeamp}, and the number of modes retained for each waveform is annotated in each panel. The other contribution to the mismatch is the slower orbital decay of eccentric binaries at fixed $p_0$, as shown in Appendix~\ref{app:morphology}.}
    \label{fig:spectrogram}
\end{figure*}

\begin{figure*}[t]
    \centering
    \begin{subfigure}[b]{0.48\textwidth}
    \centering
    \includegraphics[width=\textwidth]{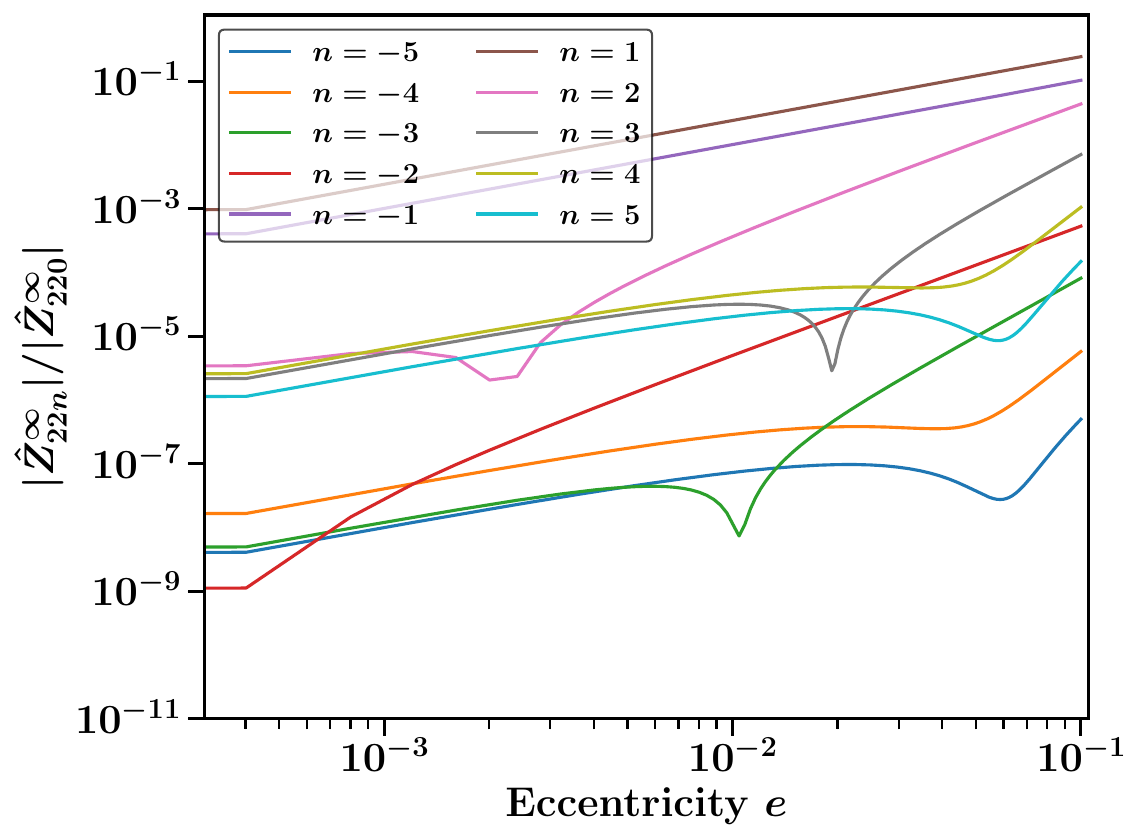}
    \caption{}
    \label{fig:modeamp_a}
    \end{subfigure}
    \hfill
    \begin{subfigure}[b]{0.48\textwidth}
    \centering
    \includegraphics[width=\textwidth]{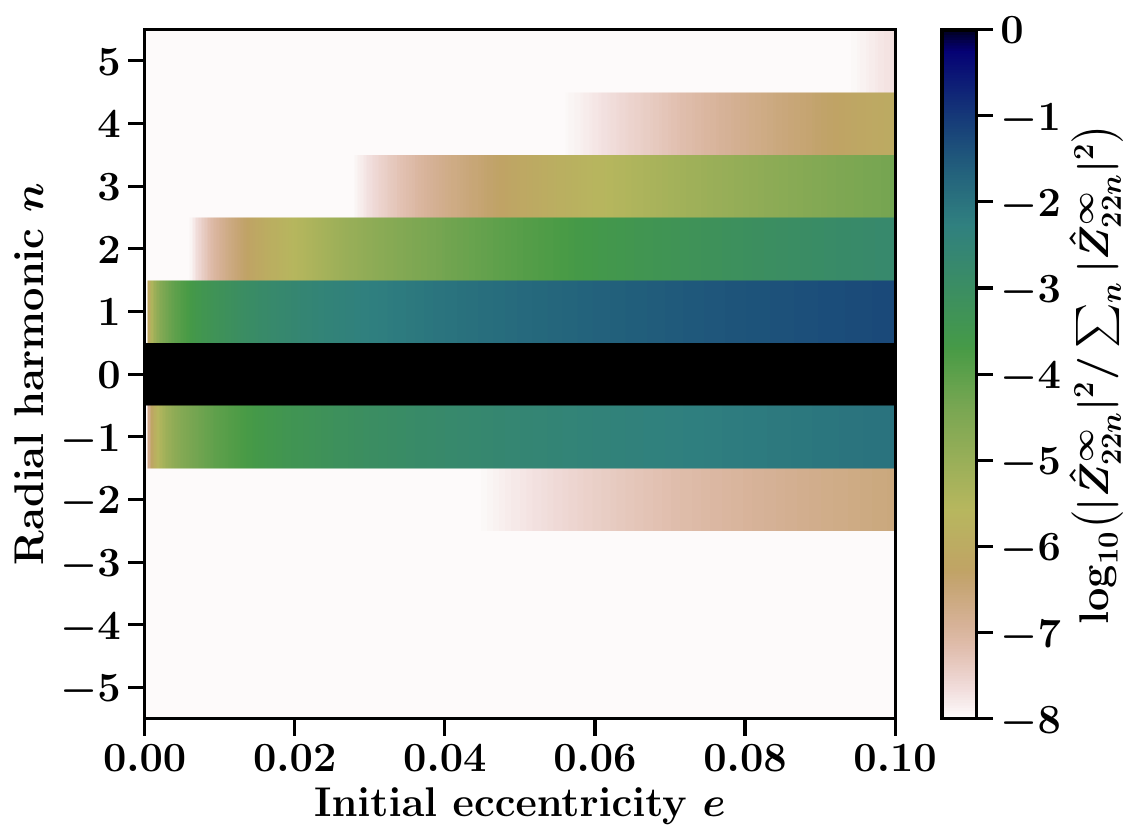}
    \caption{}
    \label{fig:modeamp_b}
    \end{subfigure}
    \caption{(a) Relative Teukolsky amplitudes of the $(2,2,n)$ radial harmonics with respect to the dominant $(2,2,0)$ mode as a function of $e_0$. (b) Normalized power carried by each $(2,2,n)$ harmonic, $P_n=|\hat Z^{\infty}_{22n}|^{2}/\sum_n|\hat Z^{\infty}_{22n}|^{2}$. Increasing eccentricity progressively redistributes power from $n=0$ into neighboring radial sidebands -- the physical origin of the spectrogram structure in Figure~\ref{fig:spectrogram} and one of the two contributions to the mismatch and parameter bias in Figures~\ref{fig:mismatch} and~\ref{fig:bias}.}
    \label{fig:modeamp}
\end{figure*}

The mismatch and parameter bias discussed above can be traced, in part, to the significant multi-harmonic nature of GWs radiated by IMRIs. In an eccentric inspiral, the Teukolsky amplitudes $\hat Z^{\infty}_{\ell mn}$ acquire non-zero contributions from radial harmonics $n\neq0$, which are strictly zero for a circular orbit. The excitation of these radial modes generates frequency sidebands spaced by the radial orbital frequency around each mode \citep{Peters:1963ux,1977ApJ...216..610T}: sidebands at $f_{mn}=mf_\varphi+nf_r$ around each pre-existing $(\ell,m,0)$ track (Figures~\ref{fig:spectrogram} and~\ref{fig:modeamp}).

Figure~\ref{fig:spectrogram} shows the resulting time-frequency spectrogram of the representative system at each of our four fiducial eccentricities. Even the circular case shows several chirping tracks, reflecting the strong mass asymmetry of an IMRI, as is generic for this class of source~\citep{Giri:2025pkx,Islam:2021zee}. As $e_0$ increases, clusters of closely spaced radial sidebands grow around every track, raising the number of waveform modes needed for an accurate reconstruction from 13 in the circular case to 31 at $e_0=0.1$ (the mode count is annotated in each panel) as shown in Figure \ref{fig:modeamp}.

Figure~\ref{fig:modeamp} quantifies the excitation for the dominant $(\ell,m)=(2,2)$ harmonic. The $n=\pm1$ sidebands are excited first and remain the leading sub-dominant contribution over the range studied, and larger $|n|$ enter as $e_0$ grows, as expected for amplitudes scaling as $|\hat Z^{\infty}_{\ell mn}|=\mathcal{O}(e^{|n|})$ when the radial motion is linearized about the circular orbit. The radial phase $\Phi_r$ is well defined at $e=0$ (the epicyclic frequency satisfies $\hat\Omega_r^2=\hat\Omega_\varphi^2(1-6/p)$, which is positive for $p>6$), but it leaves no imprint on a circular waveform because the amplitude of the radial oscillation vanishes. Power moves from $n=0$ into $n=\pm1$ and then into larger $|n|$, although $n=0$ remains dominant even at $e_0=0.1$ (Figure~\ref{fig:modeamp_b}). The number of modes retained by the \few\ selector ($\varepsilon=10^{-3}$) rises accordingly: over the full $(M,q)$ grid with $T_{\rm obs}=10\,\mathrm{hr}$, circular signals need 13 modes, while $e_0=10^{-4}$, $10^{-2}$ and $0.1$ need up to 22, 103 and 215 modes (grid minima of 13, 17 and 29).

A circular template, by construction, contains only the $n=0$ modes. It cannot represent this sideband power at any choice of mass or mass ratio, which is precisely why maximizing the match over the intrinsic parameters only partially recovers the lost overlap. This sideband content is one of two contributions to the mismatch. The other is the slower orbital decay of eccentric binaries at fixed $p$ (Appendix~\ref{app:morphology}), which the shifts in total mass and mass ratio of Figure~\ref{fig:bias} partly absorb.

\section{Discussion and Conclusions}\label{sec:discussion}

We have shown that circular waveform templates are not adequate for searches or parameter estimation of eccentric IMRIs in the decihertz band. For our representative system, the standard $3\%$ mismatch threshold is crossed at $e_0\simeq3\times10^{-2}$, well within the range supported by the waveform model across the decihertz-relevant $(M,q)$ parameter space (Section \ref{sec:waveforms}). Dynamical formation channels, in turn, are expected to leave IMRIs with non-negligible eccentricity \citep{Wen:2002km,OLeary:2008myb,Samsing:2017xmd}. The eccentricity at which circular templates fail is therefore not a corner case, but lies within the range plausible for the IMRI population that decihertz detectors such as B-DECIGO, LGWA, and IndIGO-D are designed to detect.

The threshold relevant for parameter estimation is reached considerably earlier. Because these sources are loud ($\rho\sim\mathcal{O}(10^{3})$ for B-DECIGO and IndIGO-D, and $\rho\sim\mathcal{O}(10^{2})$ for LGWA-Si at $1\,\mathrm{Gpc}$), the mismatch exceeds the statistical distinguishability criterion at $e_0\simeq5\times10^{-4}$ (B-DECIGO and IndIGO-D) and $\simeq5\times10^{-3}$ (LGWA-Si). The former is an extrapolation below the simulated eccentricity grid (Section~\ref{sec:results_emin}) for both detectors. These thresholds lie below the eccentricity at which the $3\%$ criterion is crossed everywhere in the detectable volume, and by one to two orders of magnitude at $1\,\mathrm{Gpc}$ (Figure~\ref{fig:emin_dL}). They set the eccentricity resolution floor for our representative system, and any decihertz waveform-accuracy requirement will need to respect it.

The practical implication is that decihertz IMRI search pipelines cannot simply inherit the circular or quasi-circular template banks developed for ground-based and LISA sources. Template banks, matched-filter searches, and Bayesian parameter-estimation codes for this band will need to incorporate orbital eccentricity explicitly, either through dedicated eccentric waveform models or through search statistics designed to be robust to the radial sidebands identified in Section~\ref{sec:results}. Otherwise, a fraction of the detectable eccentric IMRI population will be missed outright, and the parameters recovered for the rest will carry systematic, eccentricity-induced biases (of $\mathcal{O}(1\%)$ in $M$ and $q$ by $e_0\sim0.05$ for our representative system with B-DECIGO and LGWA-Si, and of a few percent with IndIGO-D, as shown in Figure~\ref{fig:bias}) that will propagate into any astrophysical or cosmological inference drawn from them, including constraints on the dynamical formation channels that IMRI eccentricity is meant to probe in the first place.

This paper has focused on non-spinning, equatorial inspirals within the adiabatic approximation, and only for a single representative mass and mass ratio. The analysis of the $(M,q)$ parameter space in Section~\ref{sec:waveforms} and of the mode content over the full $(M,q)$ grid in Section~\ref{sec:results_mechanism} indicates that the underlying mechanism, and the related parameter space over which it operates, are not artifacts of this particular choice. Extending the calculation to spinning and inclined systems, to the full $(M,q)$ grid, to longer observation baselines, and to full Bayesian parameter estimation with eccentric waveform models are natural next steps. These are essential to translate the qualitative conclusion of this work (that circular templates are not enough) into the quantitative template-bank and search design that DECIGO-, LGWA-, and IndIGO-D-era IMRI astronomy will require.

Before concluding, we should touch upon a few points related to the adiabatic treatment in the present context. At adiabatic order, the mass ratio enters the trajectory only as an overall rate factor while the driving fluxes are those of the $1/q \to 0$ limit and carry no residual mass-ratio dependence. The first post-adiabatic (1PA) terms, which are omitted here, contribute $\mathcal{O}(q^{0})$ radians to the accumulated phase, and therefore, they do not diminish as the binary is made more asymmetric~\citep{Burke:2023lno}. Moreover, to the best of our knowledge, no 1PA model is presently available for \textit{eccentric} orbits in the IMRI limit, and developing one is beyond the scope of the current work. Note that in the $e_0 \to 0$ limit, the eccentric signal converges to a circular waveform -- the same family as the template. Therefore, any omitted 1PA correction, being common to both signal and the template, cancel each other in the mismatch calculation. However, the same is not true for a finite eccentricity, where the eccentricity dependent 1PA contribution to the mismatch could survive. Quantifying the residual systematic would require an eccentric waveform that retains explicit mass-ratio dependence beyond an overall scaling. Such a model has recently become available in the form of a hybrid combining post-Newtonian results, which are exact in the symmetric mass ratio through 3PN, with BH perturbation theory results through 5PN and $\mathcal{O}(e^{10})$~\citep{Laxman:2026xts}. A systematic comparison against it is left to future work.

With these caveats in mind, the significance of our results goes beyond waveform modelling. For the total masses considered here, the inspiral ends at a GW frequency of at most about 1 Hz, where ground-based detectors, including the planned Einstein Telescope and Cosmic Explorer, lose sensitivity steeply. The late inspiral of these binaries, and the eccentricity it carries, is therefore out of reach of present and future ground-based observatories. They can contribute only through multiband observations, for example, by catching the merger of the lighter systems. A decihertz detector can tell an eccentric IMRI at 1 Gpc from a circular one at eccentricities as small as $5 \times 10^{-4}$ (B-DECIGO and IndIGO-D) and $5 \times 10^{-3}$ (LGWA-Si) for our representative system, one to two orders of magnitude below the eccentricity at which circular templates lose 3\% of the SNR. This opens the way to separating isolated from dynamical IMRI formation, but only if searches and parameter estimation are built on eccentric waveform models. Circular templates would hide the very eccentricity that reveals how these binaries formed.

\begin{acknowledgments}
We acknowledge the use of the IUCAA LDG computing cluster \texttt{Sarathi} for carrying out the computational and numerical work presented in this study. DG acknowledges the $33^{rd}$ meeting of the Indian Association for General Relativity and Gravitation (IAGRG), held at BITS Pilani (Pilani), where this project was initiated, and \textit{The Future of Gravitational-Wave Astronomy 2025} (code: ICTS/FGWA2025/10), organized by the International Centre for Theoretical Sciences (ICTS), where part of this work was carried out. S. Mukherjee is thankful to the Inspire Faculty Grant (DST/INSPIRE/04/2020/001332) from DST, Govt. of India, Prime Minister Early Career Research Grant (ANRF/ECRG/2024/004108/PMS) by ANRF, Govt. of India, and the New Faculty Seed Grant (NFSG/PIL/2023/P3794) provided by BITS Pilani (Pilani), India, for financial support. He (S. Mukherjee) is also grateful to the Visiting Associateship Program at IUCAA, Pune, for academic visits where a part of this work was carried out.
\end{acknowledgments}
\appendix

\section{Waveform Morphology}\label{app:morphology}
\begin{figure*}[t]
    \centering
    \includegraphics[width=\textwidth]{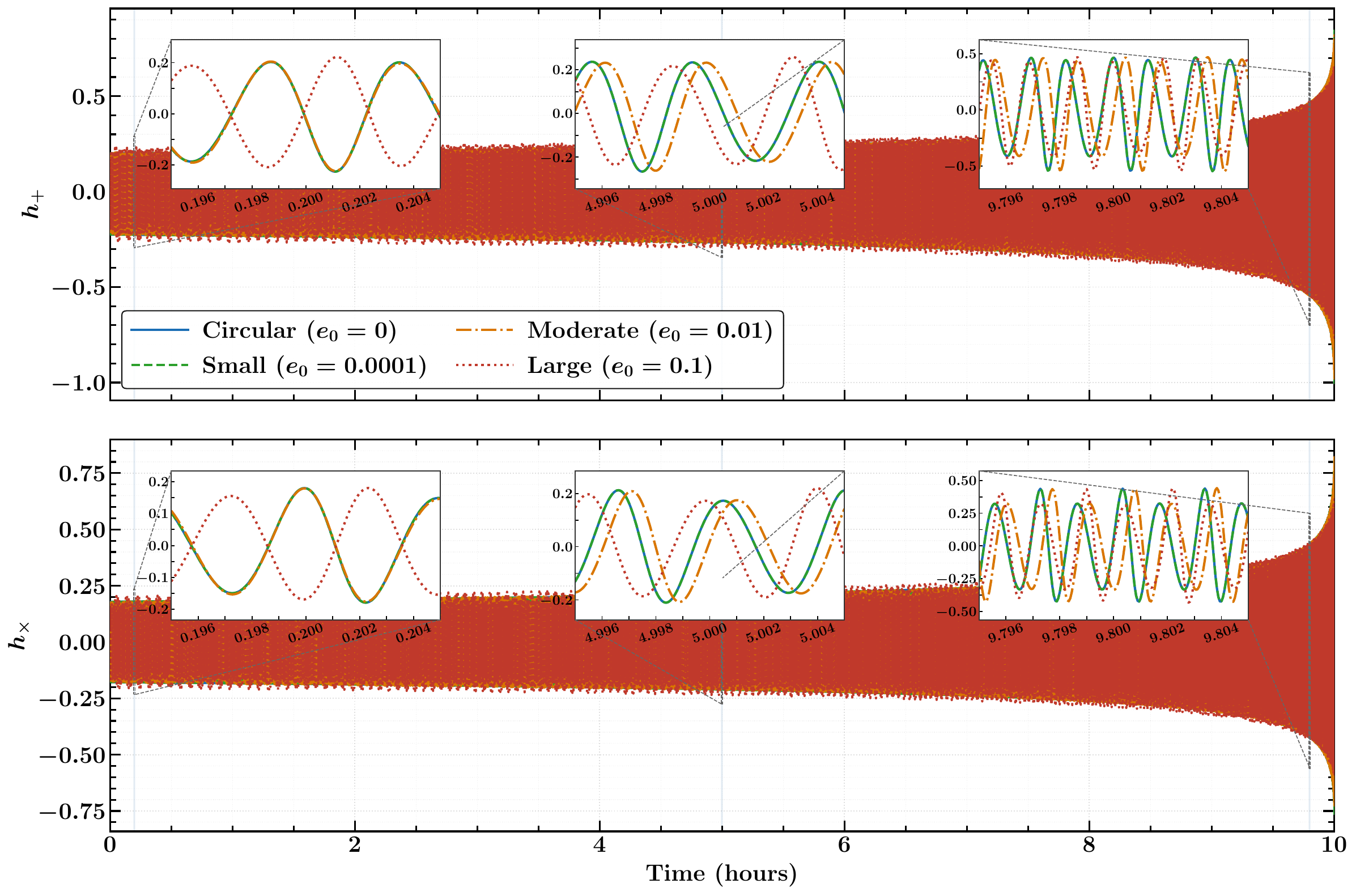}
    \caption{GW polarizations $h_+$ (upper) and $h_\times$ (lower) for the representative system at the four fiducial eccentricities. Zoom insets at the start, middle, and end of the inspiral (centred at $\approx0.2$, $5.0$, and $9.8\,\mathrm{hr}$) show the progressive dephasing and amplitude modulation introduced by eccentricity as the system approaches the ISCO.}
    \label{fig:waveform_td}
\end{figure*}

\begin{figure*}[t]
    \centering
    \includegraphics[width=\textwidth]{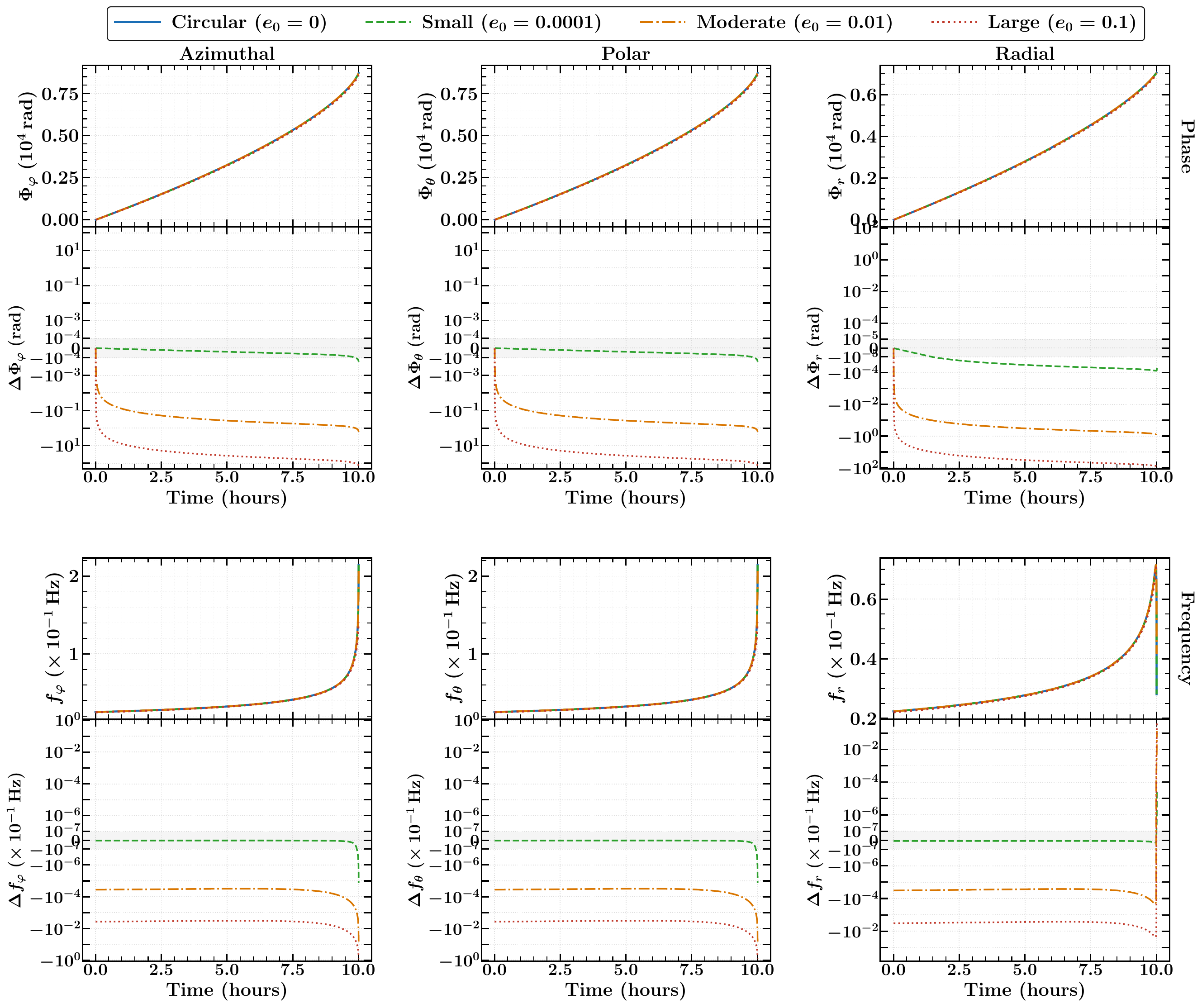}
    \caption{Cumulative phases (upper blocks) and instantaneous frequencies (lower blocks) for the azimuthal (left), polar (middle), and radial (right) degrees of freedom, for all four fiducial eccentricities. The residual panels immediately below each block show the difference relative to the circular case on a symmetric-log scale. For Schwarzschild equatorial orbits $\Phi_\theta\equiv\Phi_\varphi$ (Section~\ref{sec:waveforms}), so the middle column repeats the left one.}
    \label{fig:phasefreq}
\end{figure*}

In Figure~\ref{fig:waveform_td}, we compare the time-domain polarizations $h_+$ and $h_\times$ for the representative system at all four fiducial eccentricities, computed at sky position $(\theta,\varphi)=(\pi/3,0)$ and luminosity distance $D_L=1\,\mathrm{Gpc}$. For $e_0=10^{-4}$, the eccentric and circular waveforms are nearly indistinguishable over most of the $10\,\mathrm{hr}$ inspiral, since sub-dominant harmonic amplitudes scale as positive powers of $e$ (Section~\ref{sec:results_mechanism}). However, the small initial frequency offset accumulates coherently over the inspiral: for $e_0=0.1$ it reaches $\approx43$ cycles of the dominant $m=2$ harmonic by the end of the $10\,\mathrm{hr}$ window. Two effects separate the eccentric waveform from the circular one. The first is a phase offset: at the same initial semi-latus rectum $p_0$, an eccentric orbit has a lower orbital frequency, and although it radiates more energy, its orbital decay at fixed $p$ is slower, so it falls progressively further behind (Figure~\ref{fig:phasefreq}). The offset is already visible in Figure~\ref{fig:waveform_td} at $\approx0.2\,\mathrm{hr}$, where the $e_0=0.1$ waveform is about half a cycle out of phase with the circular one (first inset). The second, independent of the first, is that the excitation of radial harmonics imprints an amplitude modulation absent from the circular waveform.

Figure~\ref{fig:phasefreq} shows the cumulative azimuthal, polar, and radial phases and their instantaneous frequencies for all four eccentricities. The azimuthal phase accumulates to $\Phi_\varphi\gtrsim9\times10^{3}\,\mathrm{rad}$ over the final ten hours (more than $10^{3}$ orbital cycles), so even minute secular changes in the orbital evolution matter for matched filtering. The residual phase relative to the circular case, $\Delta\Phi_\varphi=\Phi_\varphi(e_0)-\Phi_\varphi(0)$, is negative throughout the inspiral for $e_0=10^{-2}$ and $0.1$, and of order $10^{-4}\,\mathrm{rad}$ or smaller for $e_0=10^{-4}$.

A negative $\Delta\Phi_\varphi$ means that the eccentric binary lags the circular one. The frequency residual in the lower panels is negative at all times for all three eccentricities, so the eccentric binary always has the lower orbital frequency. Two effects contribute, and both act in the same direction. First, at $t=0$ both orbits share $p_0$, so the initial offset is purely kinematic. To $\mathcal{O}(e^2)$, the Schwarzschild azimuthal frequency at fixed $p$ is~\citep[using Eq. 2.15, 2.16, and 2.37]{Cutler:1994pb}
\begin{equation}
  \frac{\Delta f_\varphi}{f_\varphi} = C(p)\,e^2, \qquad
  C(p) = -\frac{3(p^2-10p+22)}{2(p-2)(p-6)}
       \;\xrightarrow{\,p\to\infty\,}\; -\frac{3}{2},
\end{equation}
which recovers the Newtonian factor $(1-e^2)^{3/2}$ at large $p$ and is valid for $e\ll p-6$. For $p_0=25.04$, $C(p_0)=-1.363$, about $9\%$ below the Newtonian value. The trajectories give $(\Delta f_\varphi/f_\varphi)/e_0^2=-1.36296$, $-1.36291$ and $-1.36109$ at $t=0$ for $e_0=10^{-4}$, $10^{-2}$ and $0.1$; the first agrees with $C(p_0)$ to $2\times10^{-5}$, and the departures at larger $e_0$ are the expected $\mathcal{O}(e_0^2)$ corrections. Second, although an eccentric orbit radiates more energy than a circular one at the same $p$, it radiates less angular momentum, so its orbital decay at fixed $p$ is slower; at leading post-Newtonian order $\dot p\propto -p^{-3}(1-e^2)^{3/2}(1+\tfrac{7}{8}e^2)\simeq -p^{-3}(1-\tfrac{5}{8}e^2)$~\citep[Eqs.~4.2--4.4]{Cutler:1994pb}. This does not contradict the familiar result that eccentric binaries merge faster~\citep{Peters:1964zz}, which compares orbits at fixed semi-major axis; at fixed $p$ the eccentric orbit has the larger $a=p/(1-e^2)$ and, although $a$ shrinks faster, it takes longer to reach plunge. The eccentric binary therefore remains at larger $p$, and its frequency deficit grows secularly even as $e$ decays: for $e_0=10^{-4}$, $|\Delta f_\varphi/f_\varphi|$ grows from $1.4\times10^{-8}$ at $t=0$ to $2.5\times10^{-4}$ at the end of the window, while $e$ falls to $4\times10^{-5}$. The resulting dephasing at the end of the window is $\Delta\Phi_\varphi=-1.7\times10^{-4}$, $-1.65$ and $-136\,\mathrm{rad}$, i.e.\ $\Delta\Phi_\varphi/e_0^2\simeq-1.7\times10^{4}$, $-1.65\times10^{4}$ and $-1.36\times10^{4}\,\mathrm{rad}$, consistent with the $\mathcal{O}(e_0^2)$ scaling expected at small eccentricity. The radial phase is integrated by the trajectory module for all $e$, including $e=0$. However, it only enters into the waveform when eccentricity excites $n\neq0$ modes, through $\Phi_{mn}=m\Phi_\varphi+n\Phi_r$. The excitation of these modes, and their resulting sideband structure, are discussed in Section~\ref{sec:results_mechanism}.

\section{Explicit Form of the Match for Multipolar Waveforms}\label{app:match}
The match entering the fitting factor of the main text is stated as the overlap maximized over a relative time and phase. For a waveform containing a single azimuthal harmonic, this phrase is unambiguous. For multipolar waveforms as in the present case, where harmonics $(\ell,m,n)$ carry distinct fundamental phases $\Phi_{mn}=m\,\Phi_\varphi+n\,\Phi_r$, it is not. In the subsections below, we discuss the statistic as computed, identifying precisely which time and phase are being maximized, and explain why this choice is adequate for the conclusions of this work.

\subsection{The adopted Statistic}\label{app:match_def}

In addition to the real noise-weighted inner product of the main text, here we define its complex counterpart -- 
\begin{align}
    \langle a\,|\,b\rangle \;\equiv\; 4\int_{f_{\rm low}}^{f_{\rm high}}
    \frac{\tilde a(f)\,\tilde b^{*}(f)}{S_n(f)}\,{\rm d}f ,
    \qquad
    (a\,|\,b)\;=\;{\rm Re}\,\langle a\,|\,b\rangle .
    \label{eq:app_complex_ip}
\end{align}
All matches in this work are computed directly on the complex strain $h=h_+-ih_\times$, evaluated at fixed source-frame viewing angles and represented by its positive-frequency content on a common frequency grid of spacing $\Delta f = 10^{-5}$\, Hz over the band $[f_{\rm low},f_{\rm high}]=[10^{-3},1]\,{\rm Hz}$.

The template is assigned the standard two-parameter extrinsic family consisting of a rigid time translation $t_0$ and a single overall phase $\Phi_0$,
\begin{align}
    \tilde h_{\rm t}(f;\,t_0,\Phi_0)
    \;=\;
    \tilde h_{\rm t}(f)\;e^{\,i\,(\Phi_0\,-\,2\pi f t_0)} ,
    \label{eq:app_extrinsic}
\end{align}
and the match is
\begin{align}
    \mathcal{M}(h_{\rm s},h_{\rm t})
    \;=\;\max_{t_0,\,\Phi_0}\;\frac{\bigl(h_{\rm s}\,\big|\,h_{\rm t}(\,\cdot\,;t_0,\Phi_0)\bigr)}{\sqrt{(h_{\rm s}|h_{\rm s})\,(h_{\rm t}|h_{\rm t})}}\;=\;\frac{\displaystyle\max_{t_0}\;\bigl|z(t_0)\bigr|}{\sigma_{\rm s}\,\sigma_{\rm t}},
    \label{eq:app_match}
\end{align}
with $\sigma^2\equiv(h\,|\,h)$ and
\begin{align}
    z(t_0)\;\equiv\;
    4\int_{f_{\rm low}}^{f_{\rm high}}
    \frac{\tilde h_{\rm s}(f)\,\tilde h_{\rm t}^{*}(f)}{S_n(f)}\;
    e^{2\pi i f t_0}\,{\rm d}f .
    \label{eq:app_z}
\end{align}
The maximization over $\Phi_0$ is analytic, $\max_{\Phi_0}{\rm Re}\bigl[e^{-i\Phi_0}z\bigr]=|z|$. The maximization over $t_0$ is performed in two stages, as implemented in the \texttt{optimized\_match} function of \textsc{PyCBC} \citep{Usman:2015kfa}. An inverse fast Fourier transform locates the peak on the discrete lag grid of spacing $\delta t = 1 / (2f_{\rm high})=0.5\,\rm{s}$, after which the template is shifted to that lag and $|z(t_0)|$ is maximized over a continuous sub-sample offset by Brent minimization, with the shift applied directly in the frequency domain. Because the time maximization is not restricted to the discrete grid, the mismatches reported here are not limited by the sampling of the correlation, and remain reliable at the $\mathcal{MM}\sim10^{-6}$ level relevant to Section~\ref{sec:results_emin}. To quantify the numerical floor, we compared the circular signal with copies of itself shifted by sub-sample times and an arbitrary phase, and with a signal generated at $e_0=10^{-6}$, whose physical mismatch with the circular signal is $\sim10^{-16}$. In all cases $1-\mathcal{M}\lesssim3\times10^{-14}$ for every noise curve, more than six orders of magnitude below the smallest distinguishability threshold of Section~\ref{sec:results_mismatch}, $\mathcal{MM}\sim1\times10^{-7}$ (IndIGO-D). The mode-selection threshold sets a separate, physical floor. At small $e_0$ the radial sidebands carry a power fraction $\propto e_0^2$, part of which falls below $\varepsilon=10^{-3}$ and is discarded. Lowering $\varepsilon$ to $10^{-5}$ for both signal and template raises the mismatch at the injected masses by $\approx(1.5$--$2)\times10^{-6}\,(e_0/10^{-3})^2$, i.e.\ by $\approx40$--$105\%$ at $e_0=10^{-3}$, by $\lesssim13\%$ at $e_0=3\times10^{-3}$, and by $\lesssim2\%$ at $e_0=10^{-2}$; the change is negligible near the $3\%$ crossing. Because the discarded sideband power has no counterpart in a circular template, it cannot be absorbed by the mass refit, and restoring it can only lower the fitting factor. The low-$e_0$ end of Figure~\ref{fig:mismatch} is therefore a lower bound on the mismatch, and the values of $e_{0,\mathrm{min}}$ in Section~\ref{sec:results_emin} are conservative upper bounds.

\subsection{On the Time and Phase}
The two maximized parameters act very differently on a multipolar waveform as compared to a single-mode waveform. The time shift $t_0$ multiplies every Fourier component by the same factor $e^{\textcolor{blue}{-}2\pi ift_0}$ and is therefore a rigid translation of the summed waveform: all $(\ell,m,n)$ modes shift together. A change of the arrival (or coalescence) time is common to all modes, so this maximization is exact for arbitrary mode content; no mode-dependent generalization of the time freedom is required. The phase $\Phi_0$ is a single constant added to the Fourier phase of the template, the same at every frequency and hence the same for every mode (that is a global rotation of the complex strain). Its physical content is as follows: Because $h=h_+-ih_\times$ carries definite spin weight $-2$, a rotation of the polarization basis about the line of sight by an angle $\psi_{\rm pol}$ acts as $h\to e^{-2i\psi_{\rm pol}}h$ on every mode simultaneously; in the idealized complex-strain response used here, $\Phi_0$-maximization is therefore \textit{exactly} the maximization over the polarization angle, for arbitrary mode content. By contrast, a shift of the template's reference orbital and radial phases acts mode-dependently. Under $\Phi_\varphi\to\Phi_\varphi+\delta\varphi$ and $\Phi_r\to\Phi_r+\delta_r$, each harmonic $h_{\ell mn}$ of the mode sum transforms as
\begin{align}
    h_{\ell mn}\;\longrightarrow\;
    h_{\ell mn}\;e^{-i\,(m\,\delta\varphi\,+\,n\,\delta_r)} .
    \label{eq:app_modephase}
\end{align}
Only when every retained mode shares a single azimuthal number $m$ (and, for eccentric content, a single $n$), Equation~\ref{eq:app_modephase} reduces to a constant rotation, and only then is the $\Phi_0$-maximization equivalent to a maximization over the orbital (coalescence) phase, with $\delta\varphi=-\Phi_0/m$; this is the familiar dominant-mode, $(\ell,|m|)=(2,2)$ case, in which ``time and phase" unambiguously means coalescence time and coalescence phase. For our multipolar waveforms, this equivalence does not hold, and Equation~\ref{eq:app_match} is the standard single-phase statistic of matched-filter searches \citep{Allen:2005fk}, not the mode-wise (``generalized'') match of the higher-multipole search literature. In the latter case, each $m$-harmonic is assigned an independent phase~\citep{Capano:2013raa,Harry:2017weg}, and Equation~\ref{eq:app_match} does not implement an exact maximization over the template's reference orbital phase $\Phi_{\varphi,0}$. Note also that our recovery templates are circular, containing only $n=0$ modes, and the template waveform is independent of the reference radial phase. So the only physical phase freedom not exactly spanned by $\Phi_0$ is the differential action of $\Phi_{\varphi,0}$ across harmonics of different $m$, i.e., the relative phasing of the subdominant ($m\neq 2$) families, which $\Phi_0$ cannot adjust independently of the dominant $m=2$ family.

\subsection{Adequacy}\label{app:match_adequacy}
Despite its limitations, the adopted maximization scheme is operationally appropriate for the present context. Equation~\ref{eq:app_match} is precisely the per-template statistic that standard matched-filter pipelines compute: an inverse FFT over arrival time and an analytic maximization over one overall phase~\citep{Allen:2005fk}. The fitting factors and mismatches we report here, therefore, measure the effectualness of a circular-template search \textit{as such searches are actually implemented}. Recently, the statistics that assign independent phases to different harmonics have been developed~\citep{Capano:2013raa,Harry:2017weg}. These are more sensitive to sources with strong higher-harmonic content, but are not the default in the current pipelines. Our use of the standard single-phase statistic therefore reflects what routine circular-template searches achieve. Because the mode-wise statistic maximizes over additional phases, it can only raise the match, so the mismatches reported here are upper bounds on those of a mode-wise search. The additional freedom concerns only the relative phasing of the $m\neq2$ families, so the difference is expected to be small as long as these carry a small fraction of the power.


\bibliography{sample701}{}
\bibliographystyle{aasjournalv7}

\end{document}